\documentclass[12pt]{elsarticle}

\makeatletter
\def\ps@pprintTitle{%
  \let\@oddhead\@empty
  \let\@evenhead\@empty
  \def\@oddfoot{\reset@font\hfil\thepage\hfil}
  \let\@evenfoot\@oddfoot
}
\makeatother

\usepackage{graphicx}
\usepackage{amssymb}
\usepackage{mathtools}
\usepackage{amsmath}
\usepackage{chemformula}
\usepackage{amsthm}
\usepackage{calrsfs}
\DeclareMathAlphabet{\pazocal}{OMS}{zplm}{m}{n} 
\usepackage [english]{babel}
\usepackage [autostyle, english = british]{csquotes}
\MakeOuterQuote{"}

\usepackage{breqn}
\usepackage{lineno}

\newdefinition{defin}{Definition}

\DeclareMathOperator{\argmin}{arg \, min} 

\newcommand{\R}{\mathbb{R}}

\newcommand{\norm}[1]{\left\lVert#1\right\rVert}
\newcommand{\xft}{H_{f_t}}

\begin{document}

\begin{frontmatter}


\title{Global Sensitivity and Domain-Selective Testing for Functional-Valued Responses: An Application to Climate Economy Models}



\author[CA,DIG,RHUL]{Matteo Fontana}
\author[DIG,EIEE]{Massimo Tavoni}
\author[MOX]{Simone Vantini}

\address[CA]{Corresponding Author, matteo.fontana@rhul.ac.uk}
\address[DIG]{Department of Management, Economics and Industrial Engineering, Politecnico di Milano, Milan, Italy}
\address[EIEE]{RFF-CMCC European Institute on Economics and the Environment, Centro Euro-Mediterraneo sui Cambiamenti Climatici, Milan, Italy}
\address[MOX]{MOX - Modelling and Scientific Computing Laboratory, Department of Mathematics, Politecnico di Milano, Italy}
\address[RHUL]{Now at Royal Holloway, University of London, United Kingdom}

\begin{abstract}

Understanding the dynamics and evolution of climate change and associated uncertainties is key for designing robust policy actions.
Computer models are key tools in this scientific effort, which have now reached a high level of sophistication and complexity. Model auditing is needed in order to better understand their results, and to deal with the fact that such models are increasingly opaque with respect to their inner workings.
Current techniques such as Global Sensitivity Analysis (GSA) are limited to dealing either with multivariate outputs, stochastic ones, or finite-change inputs. This limits their applicability to time-varying variables such as future pathways of greenhouse gases.
To provide additional semantics in the analysis of a model ensemble, we provide an extension of  GSA methodologies tackling the case of stochastic functional outputs with finite change inputs.
To deal with finite change inputs and functional outputs, we propose an extension of currently available GSA methodologies while we deal with the stochastic part by introducing a novel, domain-selective inferential technique for sensitivity indices.
Our method is explored via a simulation study that shows its robustness and efficacy in detecting sensitivity patterns. We apply it to real world data, where its capabilities can provide to practitioners and policymakers additional information about the time dynamics of sensitivity patterns, as well as information about robustness.




\end{abstract}


\end{frontmatter}


\section{Introduction}
Climate Change is a key issue for policymaking, both on a tactical as well as strategic level: according to the last Intergovernmental Panel for Climate Change (IPCC) \cite{Stocker2018IPCCGlobal} more decisive actions must be undertaken now, if we want to meet international agreements and preserve the integrity of the planet and its inhabitants.

A fundamental tool to understand and explore the complex dynamics that regulates this phenomenon is the use of computer models. In particular, the scientific community has oriented itself towards the use of coupled climate-energy-economy models, also known as Integrated Assessment Models (IAM). These are pieces of software that integrate climate, energy, land and economic modules, to generate predictions about decision variables for a given period (usually, the next century). They belong to two very different paradigms \citep[see e.g.][]{weyant_contributions_2017}: detailed process models which have provided major input to climate policy making and assessment reviews such as those of the IPCC. And benefit-cost models such as the Dynamic Integrated Climate-Economy (DICE) model \citep{nordhaus_rolling_1993}, for which the economics Nobel prize was awarded in 2018. A classic variable of interest in this kind of analysis is the level of future $CO_2$ emissions, since these directly affect climatic variables, such as global average temperature.

Predicting a quantity for the long time scales which matter for the climate is a hard task, with a great degree of uncertainty involved. Many efforts have been undertaken to model and control this and other uncertainties, such as the development of standardized scenarios of future development, called Shared Socio-economic Pathways (SSPs) \cite{ONeill2014,Riahi2017} or the use of model ensembles to tackle the issue of model uncertainty. Given also the relative opaqueness and the complexity of IAMs, post-hoc diagnostic methods have been used, for instance with the purpose of performing Global Sensitivity Analysis. In fact, GSA methods can provide fundamental information to policymakers in terms of the relevance of specific factors over model outputs \citep{marangoni_sensitivity_2017}. Moreover, the specific methodology employed in the paper \citep{borgonovo_sensitivity_2010} is able to detect both main and interaction effects with a very parsimonious experimental design, and to do so in the case of finite changes for the input variables.

Some fundamental pieces of knowledge are still missing: given a dynamic phenomenon such as the evolution of $CO_2$ emissions in time a policymaker is interested if the input of the factor varies across time, and how. Moreover, given the presence of a model ensemble, with different modelling choices, and thus different impacts of identical input factors across different models, a key information to provide to policymakers is if the evidence provided by the model ensemble is significant, in the sense that it is "higher" than the natural variability of the model ensemble. In this specific setting we do not want just to provide a "global" idea of significance, but we also want to explore the temporal sparsity of it (e.g. I would like to know if the impact of a specific input variable is significant in a given timeframe, but fails to be "detectable" in the model ensemble after a given date). Our aim in the present work is thus threefold: we want to introduce a way to express sensitivity that allows to account for time-varying impacts, and we also want to assess the significance of such sensitivities, being able to explore the presence of temporal sparsity of the significance. 

A very natural framework to tackle this specific issue is Functional Data Analysis (FDA) \citep{ramsay_functional_2005}, the branch of statistics that deals with studying data points that come in the shape of continuous functions over some kind of domain. FDA is a niche yet established area in the statistical literature, with many applied and methodological publications in all domains of knowledge, including spatial and space-time FDA \citep{delicado_statistics_2010, madonna_new_2022, king_functional_2018,nerini_cokriging_2010,nerini_cokriging_2010, giraldo_ordinary_2011}, coastal engineering \citep{otto_statistical_2021}, environmental studies \citep{ballari_spatial_2018, maranzano_adaptive_2023}, transportation science \citep{piter_helsinki_2022} and epidemiology \citep{ruiz-medina_functional_2014}.

Methodologies for GSA that are able to deal with functional outputs are present in the literature: \cite{lamboni_multivariate_2011} propose non-time-varying sensitivity indices for models with functional outputs, based on a PCA expansion of the data. This approach is thus not capable of detecting the presence of time variations in impacts, nor does it address the issue of statistical significance of impacts. \cite{gamboa_sensitivity_2014} proposes a similar approach, without specifying a fixed functional basis, and proposing an innovative functional pick-and-freeze method for estimation. \cite{francom_sensitivity_2017} instead use a bayesian framework, based on adaptive splines to extract also in this case non-time-varying indices. In all the cited works around GSA techniques for functional outputs uncertainty is not explicitly explored. A very sound framework for the GSA of stochastic models with scalar outputs is provided in \cite{antoniano-villalobos_which_2018}. 

To our knowledge, there are no methods that deal with GSA of stochastic models with functional or multivariate outputs. Moreover, none of the works related to GSA cited in this paragraph deal with finite changes. For these reasons, to provide methodologies that are able to tackle the applicative questions mentioned above, we will provide a novel vision of GSA for functional outputs and finite changes using concepts developed while working in FDA. Namely, by exploiting the similarity between the proposed Sensitivity Analysis technique for Functional-valued outputs and Functional Linear Models \cite{ramsay_functional_2005}, we use a cutting edge non-parametric testing technique for Functional-on-Scalar Linear Models, called Interval-Wise Testing \cite{pini_interval-wise_2017} to address in a statistically sound way the issue of uncertainty.

The structure of the work is the following: in Sec. \ref{sec:sensitivityforfunvariables} we provide an extension to the theory and we define a new set of Finite Change Sensitivity Indices (FCSIs) for functional-valued responses, while in Sec. \ref{sec:significance testing} we then proceed to present and develop the methodology to assess the uncertainty associated with these FCSIs.
Finally, in Sec. \ref{sec:co2emissionsasfunctionaldata} we tackle the motivating problem: moving from \cite{marangoni_sensitivity_2017}, we extend their results by providing, using the previously developed theory, an analysis of the time variability of sensitivities in time, as well as a quantification of the statistical significance and an analysis of its sparsity. Sec. \ref{sec:conclusions} concludes and devises additional research directions.

In the Supplementary Material to this paper the interested reader can find an extensive simulation study that puts the proposed indices, estimation and inference technique to the test.

\section{Global Sensitivity for Functional Outputs}
\label{sec:sensitivityforfunvariables}

The present section deals with the mathematical background needed to the decomposition of finite changes in functional variables.
Building from \cite{borgonovo_sensitivity_2010} and \cite{Rabitz1999}, let us define the input-output (I/O) relationship of a given simulation model with a real response varying over an interval as

\begin{equation}
    \label{EQ-dependencemodel}
    f:K^p\times[t_1,t_2]\rightarrow \mathbb{R}.
\end{equation}

The input space is $K^p$, defined as the hypercube $\{ \left( x_1,\ldots,x_p \right), \,0\leq x_i \leq1,\,$ $ i=1,\ldots,p \}$\footnote{The restriction on the unit hyper-cube is for convenience, and can be easily relaxed via a normalization of the input variables}. The domain of definition of the continuously-varying, real-valued response is $[t_1,t_2]=T$.
From the previous elements, we can now define, $\forall x \in K^p$
\begin{equation}
f_x(t)=f(x,t):T\rightarrow \mathbb{R}, 
\end{equation}
which is the response of the simulation model for a given $x$, and $\forall\, t \in T$
\begin{equation}
f_t(x)=f(x,t):K^p\rightarrow \mathbb{R},
\end{equation}
which is the function generating the scalar response for a given $t$. $\forall\ t \in T$, $f_t$ belongs to the linear functional space $\xft$, while $\forall\ x \in K^p$, $f_x$ belongs to another linear functional space $H_{f_x}$

Simulation situations in which such a modelling could apply are for instance \citep{vitousek_model_2023}, where the authors simulate using a numerical model coastline shapes. $T$ in this case would be represented by the extension of coastline studied, while $f_x(t)$ would be the coastline profile, generated by a specific set of inputs $x$.
Another possible application scenario could be the plate deformation experiment of \cite{francom_sensitivity_2017}, where $T$ is the linear extension of a plate, and $f_x(t)$ is its deformation induced by input settings $x$. 

Let us define a measure $\mu$ on subsets of $K^p$ so that $\left\{ K^p,\pazocal B (K^p),\mu \right\}$ is a measure space, where $\pazocal B (K^p)$ denotes the Borel $\sigma$-algebra  on $K^p$. We then consider a subspace of $\xft$ identified by all the functions integrable with respect to $\mu$. Let us restrict $\mu$ to be a product measure with unit mass and a density, so that
\begin{equation}
    \label{EQ-productmeasure1}
    d\mu(x) = d\mu(x_1,\ldots,x_p) = \prod_{i=1}^p d\mu_i(x_i),\, \int_{0}^1 d\mu_i(x_i)=1\, \forall\, i \in \{ 1,\ldots,p \},
\end{equation}
and that
\begin{equation}
    \label{EQ-productmeasure2}
    d\mu(x)=m(x)dx = \prod_{i=1}^p m_i(x_i)dx,   
\end{equation}
where $m_i(x_i)$ is the marginal density function of the input $x_i$.

We can now define on $\xft$ the inner product $\langle\cdot,\cdot\rangle$ induced by the measure $\mu$ to be
\begin{equation}
    \label{EQ-innerproduct}
    \langle h , k\rangle = \int_{K^p} h(x)k(x)d\mu(x),\quad h(x),k(x)\in \xft.
\end{equation}
The norm $\norm{\cdot}_{\xft}$ induced by the inner product \eqref{EQ-innerproduct} is defined as:
\begin{equation}
    \label{eq-norm}
    \norm{f(x)}_{\xft} = \left( \langle f , f\rangle\right) ^{1/2} = \left( \int_{K^p} f^2(x) d\mu(x) \right)^{1/2}.
\end{equation}
Two functions $h(x)$ and $k(x)$ are orthogonal when $\langle h , k \rangle = 0 $.

Following Lemma 1 of \cite{Rabitz1999} we can say that, $\forall t \in T$, $\xft$ can be decomposed as 

\begin{equation}
    \label{eq-spacedecomposition}
    \xft=\pazocal{V}_0 \oplus \sum_{i=1}^p \pazocal{V}_{i} \oplus \sum_{i<j}^p \pazocal{V}_{i j} \oplus \ldots \oplus \pazocal{V}_{1,\ldots,p},
\end{equation}

where $\oplus$ is the usual direct sum operator, and $\pazocal{V}_0,\pazocal{V}_{i},\ldots,\pazocal{V}_{1,\ldots,p} \subset \xft$ are linear subspaces defined $\forall t$ as:

\begin{equation}
    \label{eq-subspacelist}
    \begin{aligned}
        \pazocal{V}_0  \equiv & \left\{  f_{t}\in \xft:\,f_{t;0}=C_t \right \} \\
        \pazocal{V}_{i}  \equiv & \left\{ f_{t} \in \xft : f_{t,i} = f_{t,i}(x_i), \: \text{with} \: \int_{K^p}f_{t,i}(x_i)d\mu_i(x_i) = 0  \right\} \\
        \pazocal{V}_{ij}  \equiv & \left\{ f_{t} \in \xft : f_{t;i,j} = f_{t;i,j}(x_i,x_j), \: \text{with} \: \int_{K^p}f_{t,i,j}(x_i,x_j)d\mu_k(x_k) = 0  \ , k=i,j \right\} \\
        & \vdots  \\
        \pazocal{V}_{1,\ldots,p} \equiv & \bigg\{ f_{t;1,\ldots,p} \in \xft: f_{t;1,\ldots,p} = f_{t;1,\ldots,p}(x_1,\ldots,x_p), \\ \text{with} 
        & \int_{K^p}f_{t;1,\ldots,p}(x_1,\ldots,x_p; t) d\mu_k(x_k) = 0 \ k=1,\ldots,p \ \bigg\},
    \end{aligned} 
\end{equation}

where $C$ is the constant function over $K^p$. As a corollary, if $f_t\in \pazocal{L}^1[K^p]$, \cite{Rabitz1999} and \cite{Sobol2003,borgonovo_sensitivity_2010} show that, when $\xft$ is a $\pazocal{L}^1$ space and $\forall t \in T$, the following decomposition is unique:
\begin{equation}
    \label{eq-functiondecomposition}
    f_t(x)=f_{t;0}+\sum_{i=1}^p f_{t;i}(x_i) + \sum_{i<j}^p f_{t;ij}(x_i,x_j) + \ldots + f_{t;1,\ldots,p}(x_1,\ldots,x_p),
\end{equation}
where the functions are defined recursively as

\begin{equation}
    \label{eq-functionlist}
    \begin{aligned}
    f_{t;0} \equiv & \int_{K^p} f_t(x)d\mu(x) \\
    f_{t;i}(x_i) \equiv  & \int_{K^{p-1}} f_{t}(x) \prod_{k\neq i} d\mu_k(x_k) - f_{t;0} \\
    f_{t;i,j}(x_i,x_j) \equiv  & \int_{K^{p-2}} f_{t}(x) \prod_{k\neq i,j} d\mu_k(x_k) - f_{t;i}(x_i) - f_{t;j}(x_j) - f_{t;0}.\\
    \vdots & \\
    \end{aligned}
\end{equation}

As proven in \cite{Rabitz1999}, the functions listed in Eq. \ref{eq-functionlist} are orthogonal. Due to orthogonality the $f_{t;i}(x_i), i \in 1,\ldots,p$, the first order terms in Eq. \ref{eq-functiondecomposition} can be considered $\forall t$ as the individual effects of the input parameters $x_i, i \in 1,\ldots,p$, while second order terms $f_{t;i,j}(x_i,x_j)$ as interaction effects between $x_i$ and $x_j$ and so on until $f_{t;1,\ldots,p}(x_1,\ldots,x_p)$ that be considered as $p$-th and last order interaction between all the input parameters.

Such orthogonal decomposition is fundamental in ensuring that the different contributions are indeed independent.

Let us now denote $y = f_t(x)$ to be the I/O function for a given $t$, evaluated at a generic $x \in K^p $ and $y^0=f_t(x^0)$ to be the same function evaluated at a reference point $x^0=\left[x_1^0,\ldots,x_p^0 \right] \in K^p$. $y^1=f_t(x^1)$ is instead a mutated state with regards to the reference point, identified by a shift of all model parameters to a mutated state $x^1=\left[x_1^1,\ldots,x_p^1\right]\in K^p$. Moving from \cite{borgonovo_sensitivity_2010}, we can prove that the decomposition in Eq. \ref{eq-functiondecomposition} can be expressed in terms of the quantities defined before
\begin{equation}
    \label{eq-deltadecomposition}
    \Delta y  = f_t(x^1)-f_t(x^0) = \sum_{i=1}^p \Delta f_{t;i} + \sum_{i<j} \Delta f_{t;i,j} +\ldots+\Delta f_{t;1,\ldots,p}.
\end{equation}

The calculation of the deltas defined above becomes straightforward from a computational point of view if we begin to consider as $\mu$ the Dirac-$\delta$ measure $d\mu=\prod_{i=1}^n \delta(h_i)dx_i$. In this finite-change scenario, the discrete differences in Eq. \ref{eq-deltadecomposition} can be expressed as

\begin{equation}
    \label{eq-deltacalculation}
    \begin{aligned}
    \Delta f_{t;i} =&  f_t(x_i) - f_t(x^0)\\
    \Delta f_{t;i,j} =&  f_t(x_{i,j}) - \Delta f_{t;i} - \Delta f_{t;j} - f_t(x^0), \\
    \ldots
    \end{aligned}
\end{equation}
where $f_t(x_i)=f_t(x_1^0,\ldots,x_i^1,\ldots,x_p^0)$ and $f_t(x_{i,j})=f(x_1^0,\ldots,x_i^1,\ldots,x_j^1\ldots,x_p^0)$ and so on. From a statistical point of view, the decomposition and expressions in the previous equations appear immediately as the way of analyzing a factorial experiment, where, $\forall t$ $\Delta f_{t;i}$ is the main effect of input $i$, $\Delta f_{t;i,j}$ the interactions between inputs $i$ and $j$ and so on for higher order interactions.

Apart from generating a significant simplification in estimation, restricting ourselves to discrete variations allows us, from a modelling perspective, to deal with scenarios that we want to explore, that may be represented by a moltitude of different modelling choices and settings in a model, such as the different Shared Socio-Economic Pathways in \cite{marangoni_sensitivity_2017}. Moreover, such setting could extend to any situation where a modeller would like to analyse the impact of either a discrete variation of the level of a continuous parameter, or when a categorical set of parameters is used.

For each $t\in T$, the generic $\Delta f_{t,\ldots}$ is a scalar value. We can so define
\begin{equation}
\label{eq_functionalDelta}
    \Delta f_{\ldots}(t): T\rightarrow\mathbb{R}, \Delta f_{\ldots}(t)\in X_{f_x},
\end{equation}
to be the continuous evaluation of the scalar $\Delta f_{t,\ldots}$ over $T$. Using this transformation, Eq. \ref{eq-deltadecomposition} can be transformed in its functional counterpart:
\begin{equation}
    \label{eq-deltadecompositionfunctional}
    \Delta y(t) = \sum_{i=1}^p \Delta f_{i}(t) + \sum_{i<j} \Delta f_{i,j}(t) +\ldots+\Delta f_{1,\ldots,p}(t),
\end{equation}
where $\Delta y(t)$ is the continuous evaluation of $\Delta y$, defined $\forall t \in T$.

Starting from Eq. \ref{eq-deltadecompositionfunctional} we can thus redefine the finite change sensitivity indices in \cite{borgonovo_sensitivity_2010} for functional-valued variables
\begin{defin}
We define the quantity
\begin{equation}
\label{EQ-individual}
    \phi(t)^{k}_{1,\ldots,k}=\Delta f_{1,\ldots,k}(t), k \in 1,\ldots,p,
\end{equation}
as the finite change sensitivity index (FCSI) for functional variables of order $k$. $\phi(t)^{k}_{1,\ldots,k}$ is the contribution to the finite change in $y(t)$ of the $k$th-order interaction between $x_{1},\ldots,x_{k}$.
The index can be also considered in its normalized version 
\begin{equation}
    \Phi(t)^{k}_{1,\ldots,k}=\frac{\Delta f_{1,\ldots,k} (t)}{\Delta y (t)},
\end{equation}
where $\Phi(t)^{k}_{1,\ldots,k}$ is the fraction of (functional) finite change associated with the $k$\textsuperscript{th}-order interaction between $x_{1},\ldots,x_{k}$
\end{defin}

\begin{defin}
The first order FCSI for input $x_i$ and a functional output is defined as
\begin{equation}
    \phi^1_i(t)=\Delta f_i(t),
\end{equation}
and, its normalized version
\begin{equation}
    \Phi^1_i(t)=\frac{\Delta f_i(t)}{\Delta y(t)}.
\end{equation}
The first order FCSI for functional outputs of variable $x_i$ is the impact of $x_i$ to the total finite change $\Delta f(t)$, while the normalized FCSI for functional outputs of variable $x_i$ is the fraction of total finite change imputable to $x_i$. 
Exploiting the functional nature of the present index, this can be calculated as
\begin{equation}
\label{eq_calculation-finite}
    \phi^1_i(t)=f(x_i,t) - f(x^0,t),
\end{equation}
where $f(\ldots,t)$ is the evaluation of $f_t(\ldots)$ for every $t \in T$.
These indices will be different from $0$ in all those time instants $t \in T$ where parameter $i$ has a direct impact over the output.
Both indices disregard higher order interactions, that will be comprehended in the next index.

\end{defin}
\begin{defin}
We define the total order FCSI of an input $x_i$ to be
\begin{equation}
    \phi^T_i(t) = \Delta f_i(t) + \sum_{i<j}\Delta f_{i,j}(t) + \ldots + \Delta f_{1,\ldots,p}(t) = \sum^p_{k=1}\sum_{i\in 1,\ldots,k}\phi^k_{1,\ldots,k}(t),
\end{equation}
and the normalized version as
\begin{equation}
    \Phi^T_i(t) =\frac{\Delta f_i(t) + \sum_{i<j}\Delta f_{i,j}(t) + \ldots + \Delta f_{1,\ldots,p}(t)}{\Delta y(t)} = 
    \frac{\sum^p_{k=1}\sum_{i\in 1,\ldots,k}\phi^k_{1,\ldots,k}(t)}{\Delta y(t)}.
\end{equation}

The index $\phi^T_i(t)$ is the total order FCSI for variable $x_i$, and is the contribution to $\Delta f(t)$ of a change in $x_i$ by itself ant together with all its interactions with the other parameters in the parameter space. $\Phi^T_i(t)$ is the corresponding fraction of the change.
\end{defin}
The total order FCSIs will be different from $0$ in all those time instants $t \in T$ where index $i$ has an impact over the output, either direct or through higher order interactions with the other parameters.

It can be also shown that the total order sensitivity indices equal to:

\begin{equation}
\label{reversefactorial}
 \phi_i^T = \Delta y(t) - \Delta y(t)_{(-i)} = f(x^1;t) - f \left(x^1_{(-i)};t \right),  
\end{equation}
where $x^1_{(-i)} = (x^1_1,\ldots,x^0_i,\ldots,x^1_k)$ is the model run in which all the parameters except the $i$-th one are shifted to the mutated state. Eq. \ref{reversefactorial} has also quite important consequences in terms of the calculation of FCSIs. As it is the case with \cite{borgonovo_sensitivity_2010}, only $2n$ runs are needed to calculate both $\phi^1_i(t)$ and $\phi^T_i(t)$

In order to appreciate the impact of the interaction terms only, we introduce the \emph{Interaction} ($\pazocal{I}$) FCSI.
\begin{defin}
We define the interaction sensitivity index for functional outputs as
\begin{equation}
    \phi_i^{\pazocal{I}}(t) = \sum_{i<j}\Delta f_{i,j}(t) + \ldots + \Delta f_{1,\ldots,p}(t),
\end{equation}
and its normalized version as 
\begin{equation}
\Phi_i^{\pazocal{I}}(t) =  \frac{\sum_{i<j}\Delta f_{i,j}(t) + \ldots + \Delta f_{1,\ldots,p}(t)}{\Delta y(t)},
\end{equation}
The interaction sensitivity index for functional outputs can be easily shown to be
\begin{equation}
    \label{eq:int_decomposition}
    \phi_i^{\pazocal{I}}(t) = \phi_i^T(t) - \phi_i^1(t).
\end{equation}
The sum of all interactions terms that involve $x_i.$ is thus represented by $\phi_i^{\pazocal{I}}(t)$.
\end{defin}

The interaction FCSIs will be different from $0$ in all those time instants $t \in T$ where parameter $i$ has an impact over the output through higher-order interactions.

The attentive eye can observe the similarity of our approach with respect to \cite{marangoni_sensitivity_2017}. The fundamental difference (and one of the key feature of our method) is that our indices are defined over $T$, meaning that they are able to provide insights about the impact of input variables all across the domain of definition of the output.

\section{Significance Testing for Sensitivity Indices}
\label{sec:significance testing}

In the presence of a I/O model whose output(s) are not intrinsically deterministic, it is of fundamental importance to compute the mean value of the sensitivity indices introduced in the previous section, and to compare their absolute or relative magnitude to the natural variability of the phenomenon, or to the uncertainty introduced with the modelling effort, to understand if the impact of a specific factor is significant or not with respect to the natural or modelling variability.

With a model output that is no longer deterministic, Eqs. \ref{eq_calculation-finite} and \ref{reversefactorial}, that are basically differences between two (now non-deterministic) model runs, should become, for $i=1,\ldots,p$: 

\begin{equation}
\label{eq_stochastic_first_order}
    \phi^1_i(t)=f(x_i,t) + \epsilon(t) - f(x^0,t) - \epsilon(t) = f(x_i,t) - f(x^0,t) + \Tilde{\epsilon}(t),
\end{equation}
and

\begin{equation}
\label{eq_stocastic_total}
 \phi_i^T = f(x^1;t) + \epsilon(t) - f \left(x^1_{(-i)};t \right) + \epsilon(t) = f(x^1;t) - f \left(x^1_{(-i)};t \right) + \Tilde{\epsilon}(t),
\end{equation}

where $f(\ldots;t)$ are deterministic, zero variance terms, and $\epsilon(t)$ are zero-mean error terms.
Under the additional assumption of $\epsilon(t)$ being i.i.d. we can say that
\begin{equation}
\label{expected_value}
\begin{aligned}
\mathbb{E}(\phi^1_i(t))= & f(x_i,t) - f(x^0,t)\\
Var(\phi^1_i(t))= & Var(\Tilde{\epsilon}(t)) = Var(2\epsilon(t))\\
\mathbb{E}(\phi_i^T)= & f(x^1;t) - f \left(x^1_{(-i)};t \right)\\
Var(\phi^T_i(t))= & Var(\Tilde{\epsilon}(t)) = Var(2\epsilon(t)).
\end{aligned}
\end{equation}

To estimate the FCSI from data, let us define a generic contrast between two model runs as $\delta y(t)$. We can so write, using Eq. \ref{eq:int_decomposition}:

\begin{equation}
    \label{eq_FANOVA}
    \delta y(t) = \phi_i^1(t) + \phi_i^{\pazocal{I}}(t) + \Tilde{\epsilon}(t),
\end{equation}

where $i=1,\ldots,p$. The individual effect of the variable is always present, while the interaction effect is available only on the $\delta y(t)$ used to calculate total effects.
Eq. \ref{eq_FANOVA} is an instance of a Functional ANOVA model of the FDA literature \cite{ramsay_functional_2005,Pini2018}.

Joint estimation and testing for the model in eq. \ref{eq_FANOVA} can be performed by using the techniques presented in \cite{Abramowicz2018}. Let us assume we have run a computer experiment composed of $N$ runs, for which we observe $\delta_n y(t), t \in T, n=1,\ldots,N$. We also need the additional condition that $X_{f_x} \subset \pazocal{C}^0[T] $. We assume the error terms $\Tilde{\epsilon}_n(t), t\in T $ to have finite total variance,

\begin{equation}
    \label{eq-errortermfeatures}
    \int_T \mathbb{E}\left[\Tilde{\epsilon}_n(t)^2\right]dt < \infty,\: \forall\, n =1,\ldots,N.
\end{equation}

The OLS estimation of the functional parameters in Eq. \ref{eq_FANOVA} is normally performed by minimizing the sum over the observation used for the estimation of the $L^2$ of the functional residuals with respect to $\phi_i^1(t), \phi_i^{\pazocal{I}(t)}, i=1,\ldots,p$, hence minimizing
\begin{equation}
\label{eq_functionalOLS}
    \sum_{n=1}^N \int_T \left( \delta y(t) - \phi_i^1(t) - \phi_i^{\pazocal{I}}(t)\right)^2 dt.
\end{equation}

Due to the interchangeability of summation and integration in Eq. \ref{eq_functionalOLS}, the minimization can be performed separately for every point of the domain, regardless of the covariance structure of $\Tilde{\epsilon}(t)$. This means that, for each time instant $t\in T$, the OLS estimate $\hat{\boldsymbol{\phi}}(t) =\left[\hat{\phi}^1_1(t),\ldots,\hat{\phi}^1_p(t),\hat{\phi}^{\pazocal{I}}_1(t),\ldots,\hat{\phi}^{\pazocal{I}}_p(t)\right]$ can be obtained as:

\begin{equation}
\hat{\boldsymbol{\phi}}(t) = \argmin_{\boldsymbol{\phi}} \sum_{n=1}^N \left( \delta y(t) - \phi_i^1(t) - \phi_i^{\pazocal{I}}(t)\right)^2.
\end{equation}

We are interested in performing tests on the (functional) coefficients of the regression model in Eq. \ref{eq_FANOVA}, and/or linear combinations of the functional coefficients. Moreover, we want to define what are the intervals in the domain in which we can reject the null hypothesis. We then want to perform hypothesis tests on linear combinations of functional coefficients of the form
\begin{equation}
    \label{eq-generichypo}
    \begin{cases}
        H_{0,C}: C \hat{\boldsymbol{\phi}}(t) = \boldsymbol{c}_0 (t),\:\forall\, t \in T \\
        H_{1,C}: C \hat{\boldsymbol{\phi}}(t) \neq \boldsymbol{c}_0 (t),\: \text{for some} \:t \in T,
    \end{cases}
\end{equation}
where $C$ is a real-valued full rank matrix in $\R^{(q\times2p)}$, $q$ denotes the number of linear hypotheses to be simultaneously tested and $\boldsymbol{c}_0 = [c_{0,1},\ldots,c_{0,q}]'$ is a vector of fixed functions in $\pazocal{C}^0[T]$.

Functional t-tests are a particular case of the statistical tests described in Eq. \ref{eq-generichypo}: for a given index, let $q=1$, $C=C_{index}\in $ $\R^{(2p)}$ with a $1$ in correspondence of the selected index, and $0$ otherwise, and $c_0(t)=0$. With this particular setting we are testing hypothesis of the type

\begin{equation}
    \label{eq-thypo}
    \begin{cases}
        H_{0,C}: \hat{\phi}(t)_i^z = 0,\: \forall\: t \in T, i\in 1,\ldots,p, z\in \left\{ 1,\pazocal{I} \right\} \\
        H_{1,C}: \hat{\phi(t)}_i^z \neq 0 ,\: \text{for some} \:t \in T,i\in 1,\ldots,p, z\in \left\{1,\pazocal{I}\right\}.
    \end{cases}
\end{equation}
In the case of rejection of the null hypothesis for a generic test, we want to select those intervals in $T$ where significant differences are detected. To do so, in theory we should perform an infinite family of tests  $\forall t_k \in T$, of the form
\begin{equation}
    \label{eq-pointwisehypo}
    \begin{cases}
        H^t_{0,C}: C \hat{\boldsymbol{\phi}}(t_k) = \boldsymbol{c}_0 (t_k)\\
        H^t_{1,C}: C \hat{\boldsymbol{\phi}}(t_k) \neq \boldsymbol{c}_0 (t_k).
    \end{cases}
\end{equation}
Instead of computing those tests, we use the Interval-Wise Testing (IWT) Procedure, presented in \cite{pini_interval-wise_2017}. The main idea of the method is to provide a control of the Interval-Wise Error Rate (IWER), i.e. for each interval $\pazocal{T}$ of the domain in which $H^t_{0,C}$ is true, the probability of $H^t_{0,C}$ is rejected in at least one point of the interval is less or equal to a given confidence level $\alpha$.
In order to perform domain selection, given any closed interval $I\subseteq T$ we will test the following.

\begin{equation}
    \label{eq-intervalhypo}
    \begin{cases}
        H^{I}_{0,C}: C \hat{\boldsymbol{\phi}}(t) = \boldsymbol{c}_0 (t),\:\forall\, t \in I \\
        H^{I}_{1,C}: C \hat{\boldsymbol{\phi}}(t) \neq \boldsymbol{c}_0 (t),\: \text{for some} \:t \in I,
    \end{cases}
\end{equation}

To test the linear hypotheses in \eqref{eq-intervalhypo}, we use the following test statistic

\begin{equation}
    T_C^{I} = \int_{I}T_C(t) dt,
\end{equation}

where 
\begin{equation}
    T_C(t)=\left( C \hat{\boldsymbol{\phi}}(t) - \boldsymbol{c}_0 (t) \right)'\left( C \hat{\boldsymbol{\phi}}(t) - \boldsymbol{c}_0 (t) \right),
\end{equation}
and $\hat{\boldsymbol{\phi}}(t)$ is the OLS estimate.

Let us denote with $p_c^I$ the p-value of test \eqref{eq-intervalhypo}. We compute it using the classical \emph(Freedman-Lane) permutation scheme \citep{freedman_nonstochastic_1983}, as shown and described in \cite{pesarin_permutation_2010}. This specific choice is the most commonly used for linear modelling.
In order to define significant intervals over the domain, we compute adjusted p-value functions. The adjusted p-value function $\Tilde{p_C}(t)$ at point $t$ testing general linear hypotheses identified by contrast $C$ is defined as
\begin{equation}
    \Tilde{p_C}(t):= \sup_{I:t\in I} p_c^I, t\in T.
\end{equation}

The thresholding of this function using a confidence level $\alpha$ yields intervals in which the IWER is controlled asymptotically.

The interpretation of the the p-value functions presented in the paper is thoroughly described in \cite{pini_interval-wise_2017}. In general, the \emph{adjusted} p-value function controls the type-I error rate in a interval-wise fashion, while the \emph{unadjusted} p-value function does it in a point-wise fashion. In other words, if one performs a thresholding of the unadjusted p-value function at a level $\alpha$, for each point of the domain where the null hypothesis holds, its probability of being selected is less or equal to $\alpha$, while for each point where the null hypothesis does not hold, its probability to be selected as significant goes to one as the sample size increases. In the adjusted case these properties are related to intervals: for each interval where the null hypothesis holds almost everywhere, its probability of being selected is less or equal to $alpha$, while for each interval where the null hypothesis does not hold almost everywhere, its probability of being selected goes to one when the sample size increases.

While not having been formalised by the original authors, one could interpret the area under the curves as an intuitive notion of effect "density", as opposed to sparsity: an input with a sparse (dense) effect will have a relatively high (low) area under the pvalue curve. This is because many parts of the p-value function will have relatively high (low) values).

\section{Application: Functional Global Sensitivity Analysis of an ensemble of Climate Economy Models}
    \label{sec:co2emissionsasfunctionaldata}
    
For this paper we focus on \ch{CO2} emissions as the main output of an ensemble of coupled climate-economy-energy models. Each model-scenario produces a vector of \ch{CO2} emissions defined from the year 2010 to 2090 at 10-years time intervals. This discretization of the output space is in any case arbitrary, since \ch{CO2} emissions do exist in every time instant in the interval $T=[2010,2090]$. A thorough description of the dataset used as a testbed for the application of the methods described before can be found in \cite{marangoni_sensitivity_2017}. This was one of the first paper to apply global sensitivity techniques to an ensemble of climate economy models, thus addressing both parametric and model uncertainty. We use the scenarios developed in \cite{marangoni_sensitivity_2017} which involve five models (IMAGE, IMACLIM, MESSAGE-GLOBIOM, TIAM-UCL and WITCH-GLOBIOM) that provide output data until the end of the interval $T$.

The dataset provides emission projections for a variety of socio-economic drivers, commonly used as the inputs of Integrated Assessment Models. These are represented by the Shared Socio-economic Pathways (SSPs, see \cite[e.g.]{riahi_shared_2017}). They have been decomposed for simplicity in $n=5$ different inputs: energy intensity $(END)$, fossil fuel availability $(FF)$, Gross Domestic Product per Capita $(GDPPC)$, low carbon technology developement $(LC)$ and population $(POP)$. These are the key variables driving \ch{CO2} emissions.

Each input is seen as a discrete variable with three different levels $\{SSP1,$ $SSP2,\,SSP3 \}.$ These levels represent the main diagnonal in a cartesian plane where on the two axes we have challenges to mitigation and adaptation to climate change. Specifically, SSP2 is a ``middle of the road'' scenario, in which the current patterns in terms of key variables are preserved along the century, while SSP1 and SSP3 are two ``mutated'' cases. SSP1 represents a ``green'' world, with higher GDP per capita, lower population growth, more energy efficient firms and consumers, a low availability of fossil fuels and high availability of low carbon technologies. SSP3, instead, is a polluted world, with high population growth, especially in developing countries, inequality in GDP per capita, higher energy intensity and lower efficiency, higher availability of fossil fuels and lower availability of low carbon technologies. A more thorough characterization of SSPs and \ch{CO2} emissions drivers can be found in \cite{marangoni_sensitivity_2017} and references therein. 

We assume $SSP2$ to be our reference level, and $SSP1,\,SSP3$ to be two different shifted levels. We thus compute two different sets of sensitivity indices: one for the contrast $SSP2 - SSP1$, the second one for $SSP2 - SSP3$. Thanks to the efficient definition of sensitivity indices, derived from \cite{borgonovo_sensitivity_2010}, we need $2(n+1)$ runs for each contrast. This means that the total number of experimental runs for each contrast is $2(n+1)$ times the number of models, yielding $60$

We can extract the functional profile underlying the discrete evaluation of \ch{CO2} emissions generated by IAMs by using a smoothing spline approach, based on the use of roughness penalties \citep[][Ch. 5]{ramsay_functional_2005}, thus obtaining a continuous functional evaluation of \ch{CO2} emissions during the whole course of the century.
In other words, being $r_i, i \in [2010,2020,\ldots,2090]$ the discrete data points generated by a run of the model ensemble for a given year, we want to fit a smooth function assuming the model $r_i = y(i) + \theta_i$ and a basis function expansion for the generic model run $x(t)$ of the form
\begin{equation*}
   x(t) +\sum^K_k c_k\phi_k(t) = \boldsymbol{c'\phi}(t)   = \boldsymbol{\phi(t)'c}.
\end{equation*}
In order to take into account the roughness of the smooth function in the fitting problem, we introduce a penalty term, represented the integrated curvature of $x(t)$ and computed as
\begin{equation}
    PEN_2 = \int\left[D^2x(t)\right]^2dt,
\end{equation}
where $D^2x(t)$ represents the second derivative of $x(t)$ with respect to $t$.
To compute the coefficients of the functional expansion we thus need to minimise the sum of squared errors
\begin{equation}
PENSSE_2 = (\boldsymbol{y} - \boldsymbol{\phi(t)'c})'(\boldsymbol{y} - \boldsymbol{\phi(t)'c}) + \lambda PEN_2. 
\end{equation}

We used $9$ cubic spline functions with equally spaced knots, placed in correspondence of the discrete data points. As shown in the literature \citep[see e.g.][Chapter 4, 5]{ramsay_functional_2005} the choice of a specific type of functional basis is essentially guided by the specific application, in terms of the spacing of the data points (equally vs unequally spaced) and the type of features (periodic or aperiodic). The aperiodicy of the emission profiles studied in this section calls for the use of penalised cubic splines. Moreover, such a choice, as shown in several works on the subject \citep{DeBoor2001, eilers_splines_2010}, is the one that minimizes the smoothing error. The weighting parameter of the roughness penalty $\lambda=100$ has been chosen via the minimisation of Generalized Cross Validation.

\begin{figure}
    \includegraphics[width=\linewidth]{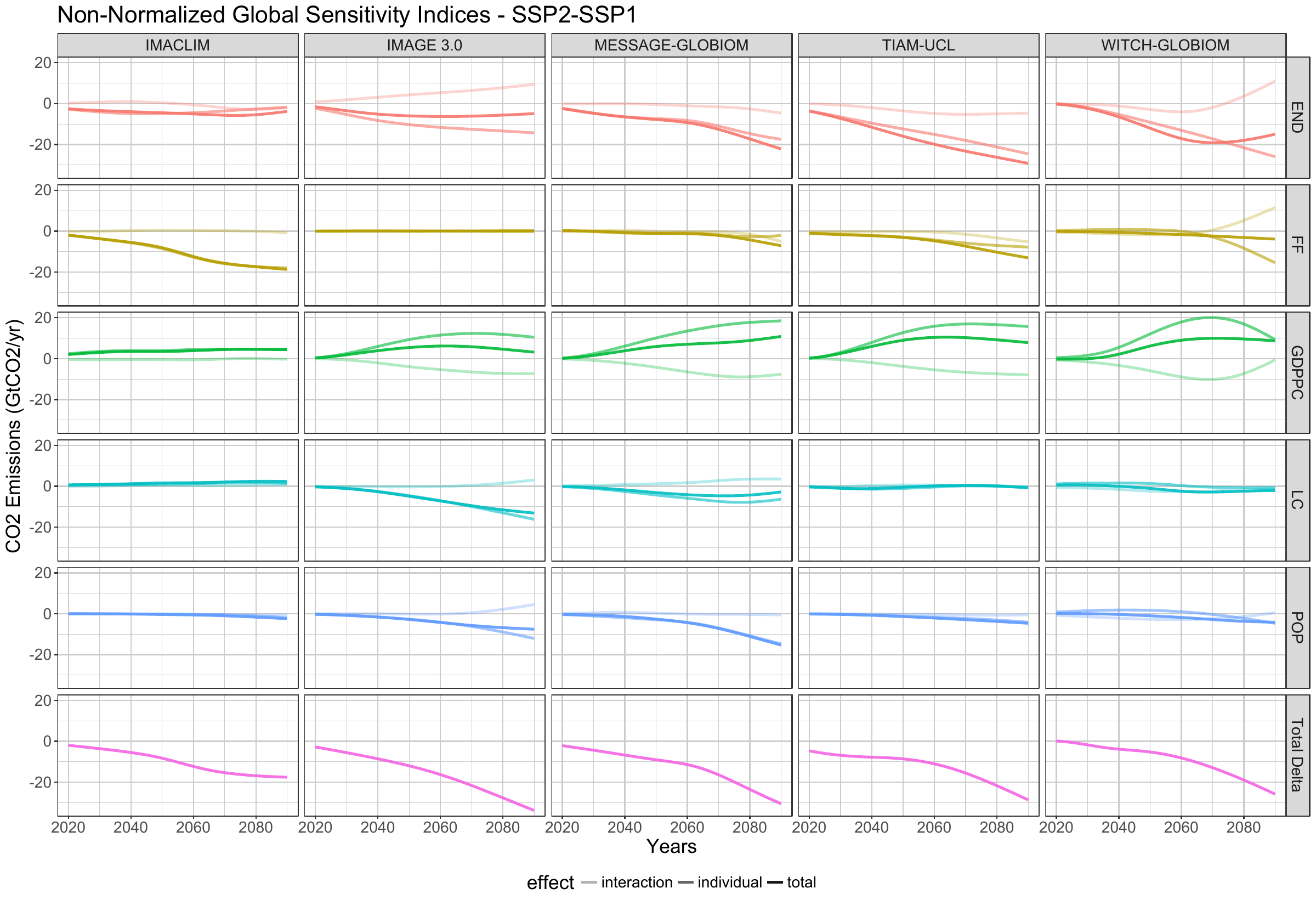}
    \caption{Functional Sensitivity indices for the SSP2 to SSP1 contrast: In all the panels, the x axis represents the time dimension (from 2020 to 2090), the y axis is the magnitude of the various non-normalized sensitivity indices, in GtCO2/year. Different rows and different colors represent different drivers, while we have different columns for different models}
    \label{fig:sensitivities2-1}
\end{figure}

\begin{figure}
    \includegraphics[width=\linewidth]{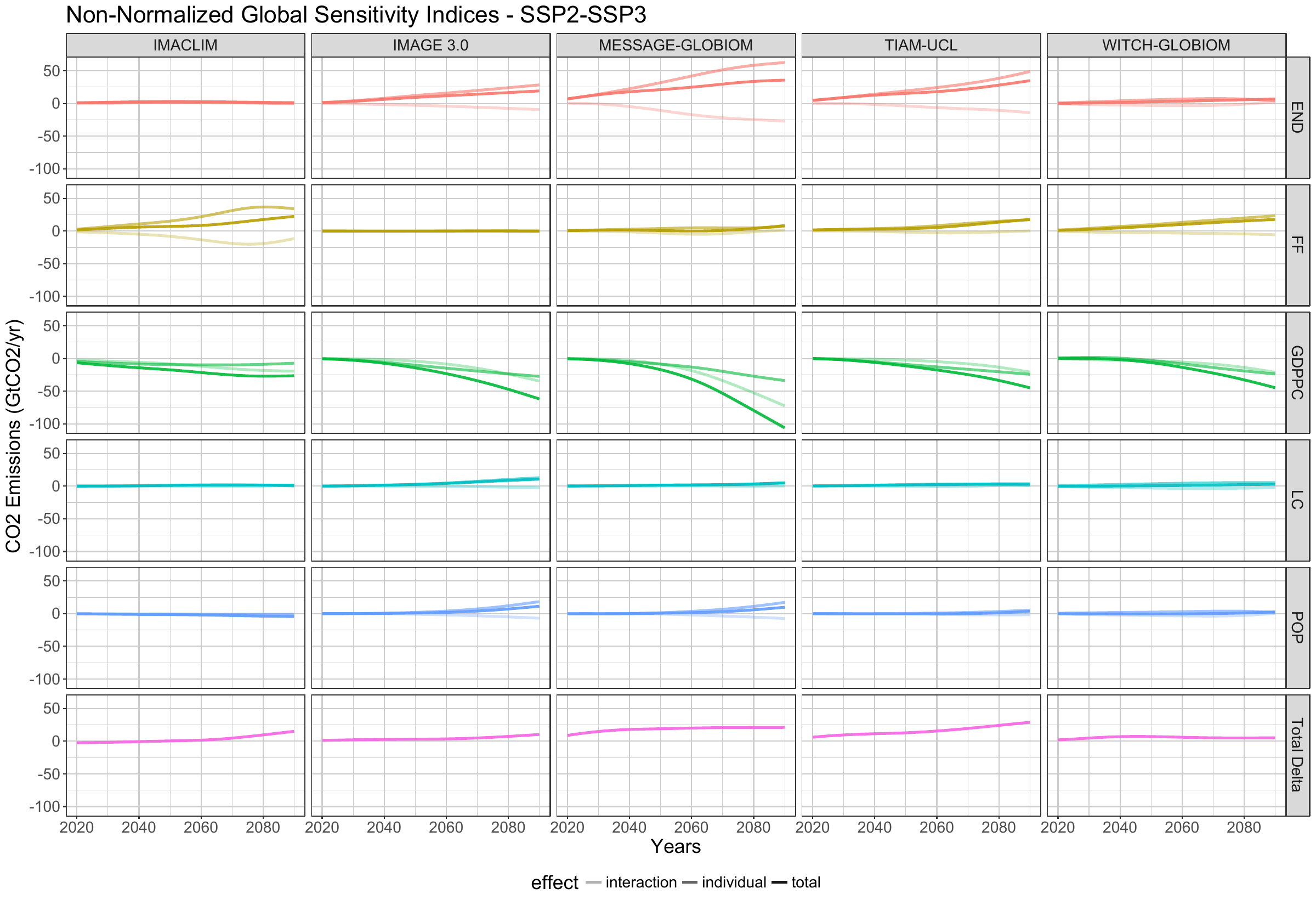}
    \caption{Functional Sensitivities for the SSP2 to SSP3 contrast. In all the panels, the x axis represents the time dimension (from 2020 to 2090), the y axis is the magnitude of the various non-normalized sensitivity indices, in GtCO2/year. Different rows and different colors represent different drivers, while we have different columns for different models}
    \label{fig:sensitivities2-3}
\end{figure}

We now have, for every combination $r$ of inputs in the dataset, a function $CO2(t)_r$ that represent the emission profile generated by the specific set of inputs. We can now calculate, for every model and experimental design, the set of $\phi^1_i,\, \phi^T_i,\, \phi^{\mathcal{I}}_i, \>l\in \left\{END,\,FF,\,GDPPC,\, LC,\,POP \right\}$ functional sensitivity indices, We restrict the domain of analysis to the interval $T=[2020,2090]$ to perform calculations only on actually generated data, and not on calibration values. The results of the sensitivity indices are shown respectively in Fig. \ref{fig:sensitivities2-1} for the SSP2 to SSP1 case, and in Fig. \ref{fig:sensitivities2-3} for the SSP2 to the SSP3 case, toghether with the total variation $\Delta CO_2(t)$.
In the SSP2 to SSP1 case we observe very similar total delta profiles on all the models, with the notable exception of the IMACLIM model, that instead shows a convex pattern.

By looking at sensitivity indices, the saliency of income levels (GDPPC) and energy intensity (END) is immediately evident, with total impacts with approximately the same magnitude and shape among models. This is in line with the findings of \cite{marangoni_sensitivity_2017}.
The sign of the total index of income (GDPPC) is positive across the entire time domain for every model: this is because SSP1 represents a generally wealthier world, where, ceteris paribus, consumption (and thus \ch{CO2} emissions) will be higher.
An interesting anomaly is represented by the sensitivity dynamics of the fossil fuels availability (FF) parameter for the WITCH-GLOBIOM model: the total effect is approximately null on $T$, while individual and interaction indices show a counteracting dynamics.

Moving to the SSP2 to SSP3 case, the time dynamics of the delta profiles is approximately similar, with the notable exception of the WITCH-GLOBIOM model that shows a very small total delta.
By looking at sensitivity indices, like in the previous case the impacts of income (GDPPC) and energy intensity (END) are the most evident. In the SSP2 to SSP3 case we also observe a probably significant time dynamics for the fossil fuel availability (FF) variable. Differently from the previous case, we also observe that the interaction effects for energy intensity (END) and income (GDPPC) have the same direction.

\begin{figure}
    \includegraphics[width=\linewidth]{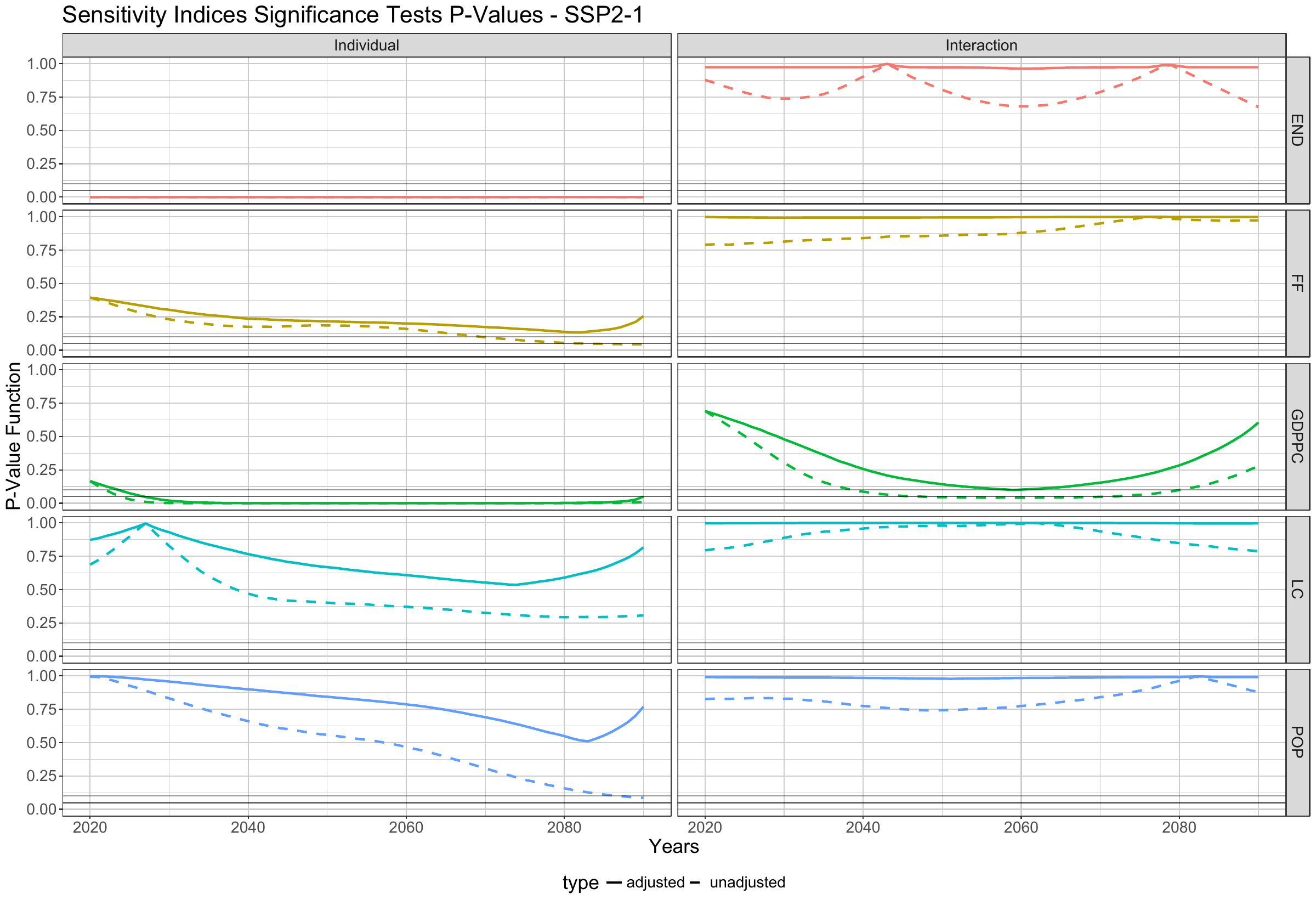}
    \caption{p-value functions for functional t-tests: SSP2 to SSP1 case. In all the panels, the x axis represents time (from 2020 to 2090), while the y are the value of the adjusted (full line) and unadjusted (dotted line) p-value functions, from 0 to 1. Rows and colors denote different drivers, while the two columns are for Individual and Interaction effects.}
    \label{fig:pval2-1}
\end{figure}

\begin{figure}
    \includegraphics[width=\linewidth]{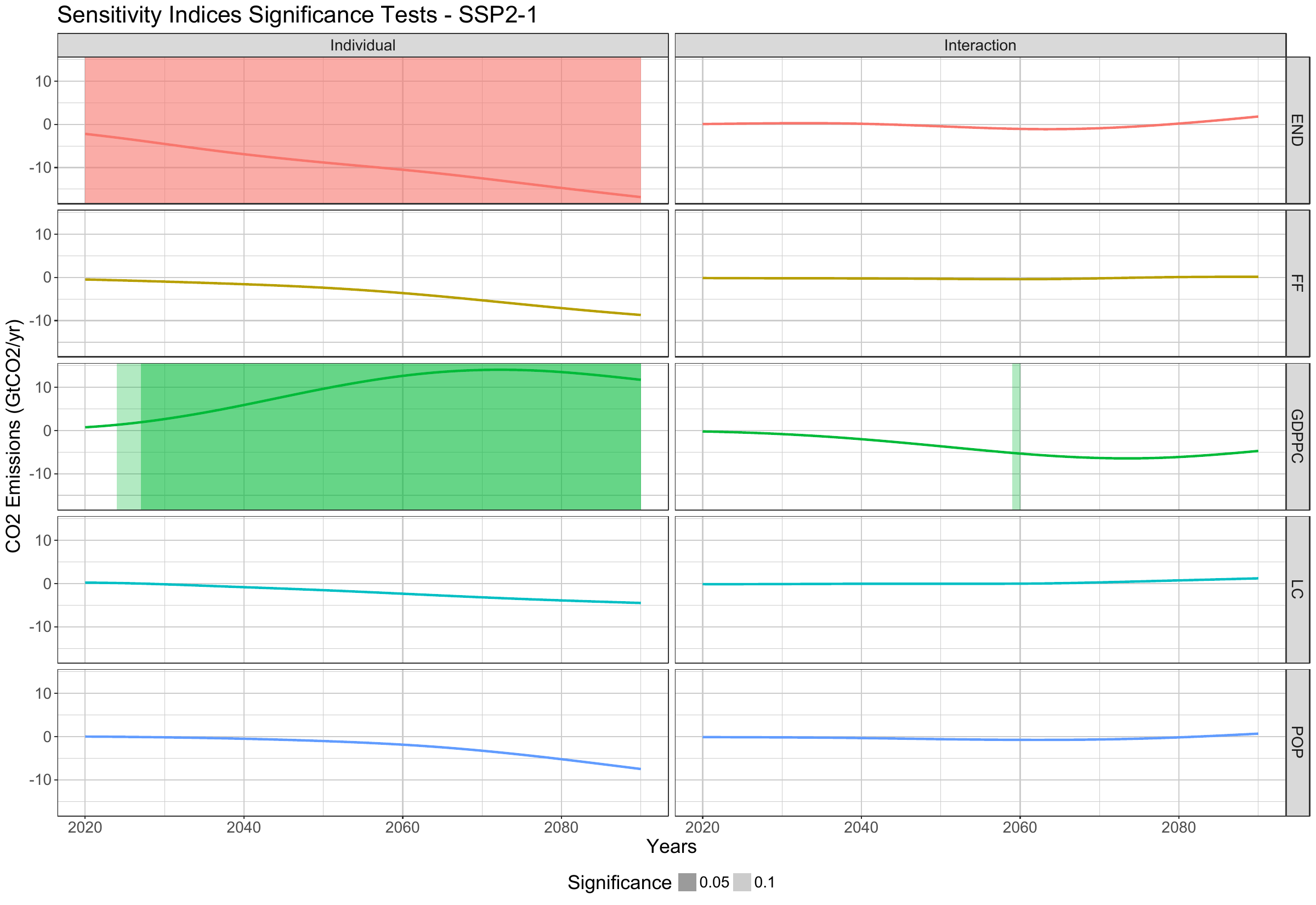}
    \caption{Functional Coefficients and Significance Levels for the SSP2 to SSP1 Case. In all the panels, the x axis represents time (from 2020 to 2090), while the y is the magnitude of the average sensitivity coefficient (in GtCO2/year), different shading levels represent different significance levels, as denoted by the adjusted p-value functions (dark shading = 0.05 significance, light shading = 0.1 significance). Rows and colors denote different drivers, while the two columns are for Individual and Interaction effects.}
    \label{fig:ttest2-1}
\end{figure}

In addition to model-specific evaluations of the functional sensitivity indices, we are also interested in testing their statistical significance. To do so, having fixed a single experimental condition $x=[x_{1},...,x_{5}]$, we consider the different runs on the 5 different models used in this experiment as replicates of the same data generating process. This is because, despite different modelling assumptions, the five IAMs are trying to capture the dynamics of the same complex phenomenon.
We can thus, following the modelling structure depicted in Eq. \ref{eq_FANOVA}, the following FANOVA model:

\begin{equation}
    \label{eq:sensitivityasregression}
    \Delta CO_2(t)_n=\hat{\phi}^1_i(t) + \hat{\phi}^{\pazocal{I}}_i(t)  + \epsilon(t)_n,\> n=1,\ldots,N,
\end{equation}
where $N=55$, $i \in \left\{END,\,FF,\,GDPPC,\, LC,\,POP \right\}$.
Estimation and domain-selective inference can be performed by using the techniques described in Sec. \ref{sec:significance testing}. We now present the results of the regressions for the SSP2 to SSP1 and the SSP2 to SSP3 scenarios in a graphical form.
The adjusted and unadjusted $P$-value functions for the $t$-tests and the functional coefficients, together with a synthetic representation of the significance level for the SSP2 to SSP1 case can be found respectively in Fig. \ref{fig:pval2-1} and \ref{fig:ttest2-1}.

After looking at the $P$-value functions, it is immediately evident how the two main significant factors are energy intensity (END) and  per capita income levels (GDPPC): their $P$-value has very low values on the whole domain of definition. FF has low values of the p-value function, but they are not low enough to render it significant over any part of the domain at a $0.05$ level. Interaction terms are not significant with the exclusion of income levels (GDPPC) in a very small and negligible part of the domain. In any case, income levels (GDPPC) Interaction p-value function values are globally very low: a slight decrease in model variability will probably render them significant. When looking at the functional coefficients, we see that the END individual mean functional sensitivity coefficient is monotonically decreasing in a linear way, while the Individual income level (GDPPC) index has instead a concave shape, with a decrease in sensitivity after the year 2070. A single year is significant in the income level (GDPPC) Interaction effect (2059).

\begin{figure}
    \includegraphics[width=\linewidth]{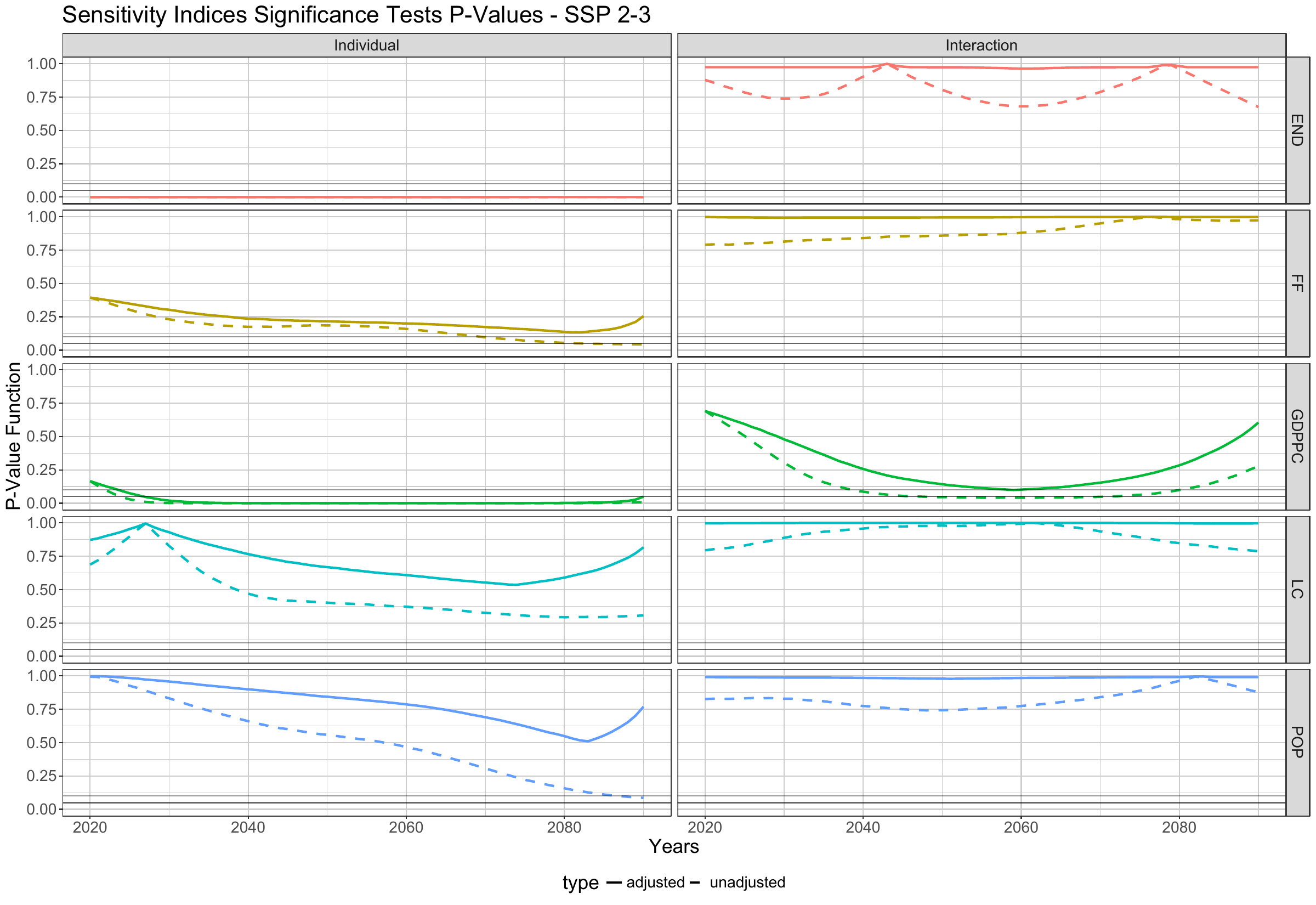}
    \caption{Pvalues for the SSP2 - SSP3 Transition. In all the panels, the x axis represents time (from 2020 to 2090), while the y are the value of the adjusted (full line) and unadjusted (dotted line) p-value functions, from 0 to 1. Rows and colors denote different drivers, while the two columns are for Individual and Interaction effects.}
    \label{fig:pval2-3}
\end{figure}

\begin{figure}
    \includegraphics[width=\linewidth]{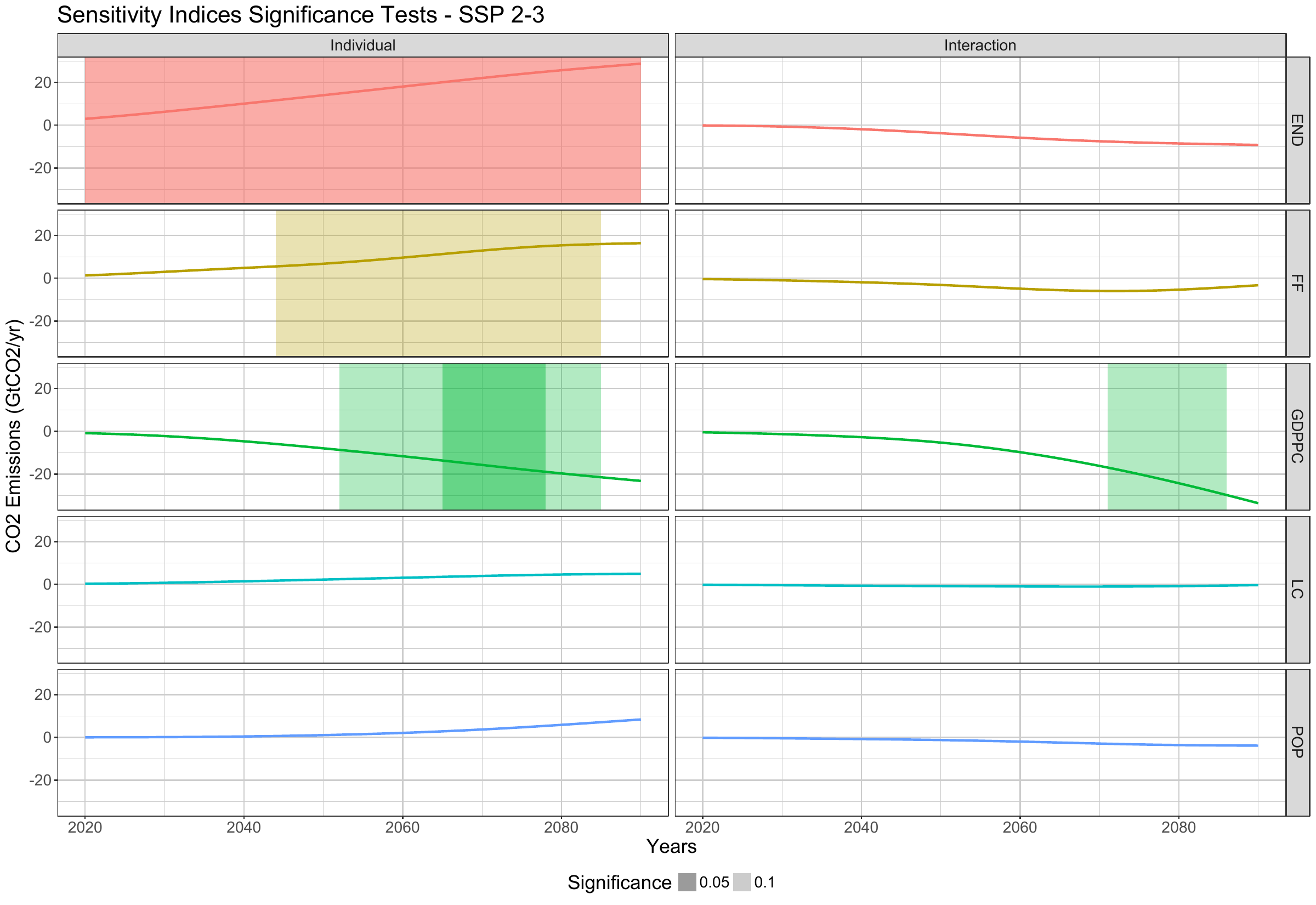}
    \caption{t-tests for the SSP2 - SSP3 Transition. In all the panels, the x axis represents time (from 2020 to 2090), while the y is the magnitude of the average sensitivity coefficient (in GtCO2/year), different shading levels represent different significance levels, as denoted by the adjusted p-value functions (dark shading = 0.05 significance, light shading = 0.1 significance). Rows and colors denote different drivers, while the two columns are for Individual and Interaction effects.}
    \label{fig:ttest2-3}
\end{figure}

The SSP2 to SSP3 case is represented in Fig. \ref{fig:pval2-3} and Fig. \ref{fig:ttest2-3}. They show similar patterns of parameters' relevance as above, but with a cleaerer time-dependent pattern. This allows to define which drivers matter the most depending on when in the future.

\section{Conclusion}
\label{sec:conclusions}
In this paper we provide a time-varying global sensitivity analysis of an ensemble of integrated assessment models used in climate change economics and modelling. We also provide an assessment of statistical significance of such sensitivities. Our work unveils novel insight inside the dynamics of CO2 emissions and relative sensitivity measures, yielding non-linearities, non-monotonicities and, generally, a behaviour that could not be captured by standard univariate global sensitivity measures.

The testing effort provides even more interesting results, showing differences between the two contrasts analyzed in this paper, and, in general, defining a sparsity in effects: The only significant factors in determining $CO_2$ emissions seem to be GDP per capita and energy intensity improvements, with fossil fuel availability being significant only in the contrast between the middle-of-the-road scenario and $SSP3$. There is no statistical evidence to affirm that interaction terms are significant, with the only notable ``near-miss'' of the interactions that involve GDP per capita. This is probably due to the pervasiveness and centrality of gross domestic product as the main economic variable inside climate-economy models. However, the importance of these drivers in determining the future climate varies over time; this allows analysts to define future periods when certain factors will be more or less relevant.

These interesting findings were rendered possible by an original extension of the current state of the art in the GSA literature, namely in the direction of defining sensitivity indices for complex data, and the statistical assessment of uncertainty on GSA indices. We prove the mathematical properties of such method, and, by exploiting the similarities between the proposed output decomposition and Functional Linear Models, we propose a novel way to perform testing over (functional) sensitivity indices.

Our findings provide a very strong signal to the climate-energy-economy modeling community that either the Shared Socio-economic Pathways are too refined to be actually significant inside a representative ensemble of models, or that, while preserving their own individuality and peculiarities in the modelling approach, that IAMs need to converge towards more homogeneous predictions.

The flexibility of the functional approach allows relatively easily to extend the proposed GSA methodology to higher dimensional functional outputs. Examples in the environmental domain could be the spatial distribution of pollutants in time \citep{thunis_sensitivity_2021}, which can be seen as a $f_x(s,t), s\in\mathbb{R}^2,t\in\mathbb{R}$, flood models \citep{brunner_challenges_2021}, where the outcome variable is the water level $f_x(s), s\in\mathbb{R}^2$, or wind forecasting \citep{pinson_verification_2012}, a situation that falls in either of the two situations described previously (simulation models of funcitons space-time , or functions for spatial data).

Such extension should be straightforward, in fact the flexibility of the interval-wise testing approach, and its subsequent extensions to a dependent data setting \citep{rimalova_inference_2022} and functional surfaces/volumes \citep[see e.g.]{lundtorp_olsen_local_2023}, allow to solve many inferential problems in geostatistical modelling, such as the ones discussed in the introduction, where I may be interested in performing domain-selective significance tests on model coefficients. Moreover, since my FCSI in this space-time case would be space-time objects, such methodology would prove fundamental in the domain selective testing for this proposed extension of our GSA framework.

Moreover, even if our GSA methodology was born in order to deal with simulation models, one could think about using it as a method to deal with Machine Learning-oriented methods dealing with functional data \citep{wang_conformal_2023}. Its role in this context would be to provide a simple yet probabilistically sound way to perform significance testing of input parameters.

\section{Acknowledgments}
Matteo Fontana acknowledges financial support from the European Research Council, ERC grant agreement no. 336155 - project COBHAM  'The role of consumer behaviour and heterogeneity in the integrated assessment of energy and climate policies'. Massimo Tavoni acknowledges financial support from the European Research Council, ERC grant agreement no. 101044703  - project EUNICE. The authors would also like to thank three anonymous reviewers for the insightful comments provided.

\section{Bibliography}

\bibliographystyle{apa}
\bibliography{references,references_old}

\end{document}


\begin{frontmatter}


\title{Supplementary Material to "Global Sensitivity and Domain-Selective Testing for Functional-Valued Responses: An Application to Climate Economy Models"}



\author[CA,DIG,RHUL]{Matteo Fontana}
\author[DIG,EIEE]{Massimo Tavoni}
\author[MOX]{Simone Vantini}

\address[CA]{Corresponding Author, matteo.fontana@rhul.ac.uk}
\address[DIG]{Department of Management, Economics and Industrial Engineering, Politecnico di Milano, Milan, Italy}
\address[EIEE]{RFF-CMCC European Institute on Economics and the Environment, Centro Euro-Mediterraneo sui Cambiamenti Climatici, Milan, Italy}
\address[MOX]{MOX - Modelling and Scientific Computing Laboratory, Department of Mathematics, Politecnico di Milano, Italy}
\address[RHUL]{Now at Royal Holloway, University of London, United Kingdom}
\end{frontmatter}



\section{Simulation Study}
\begin{figure}
    \includegraphics[width=\linewidth]{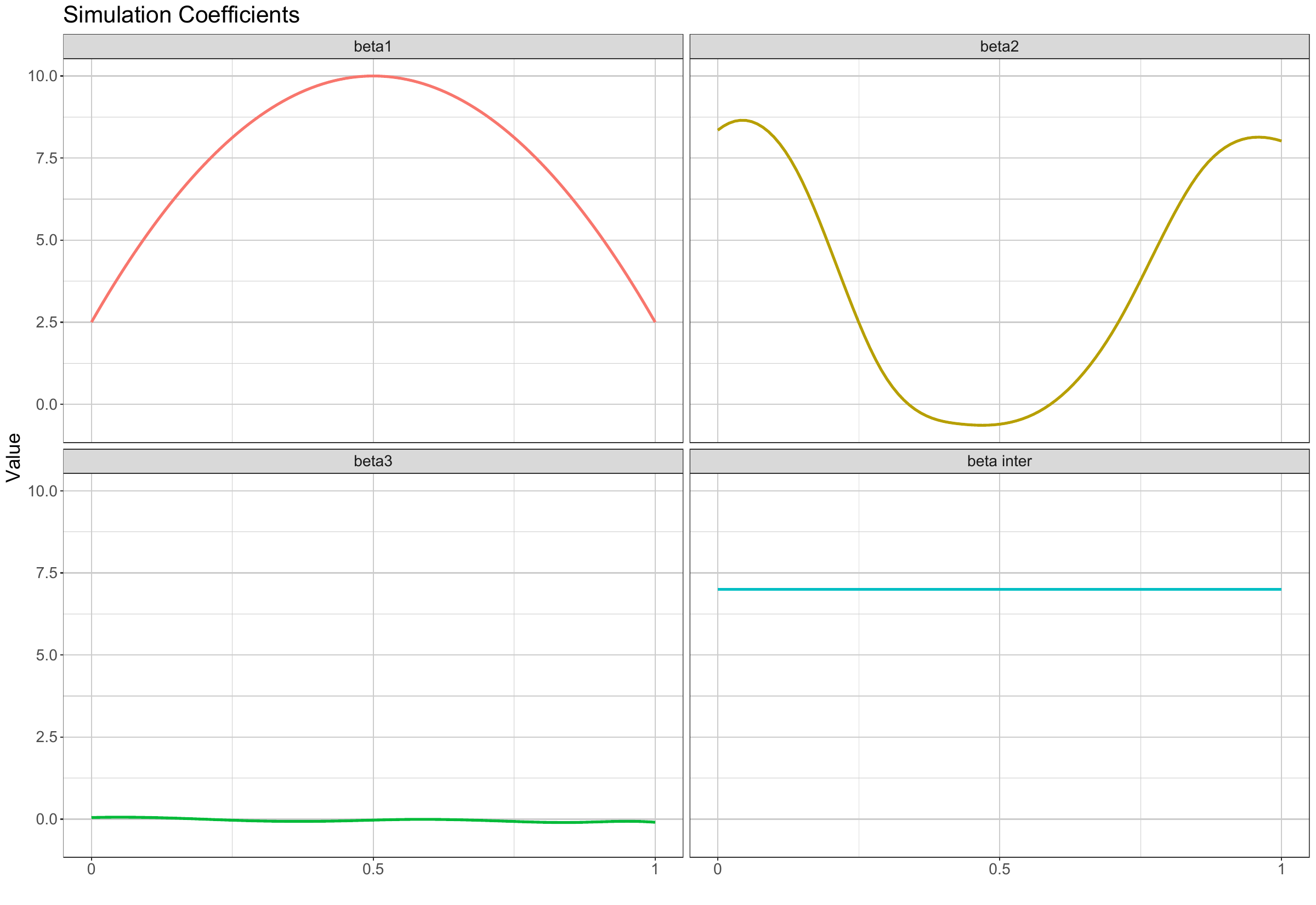}
    \caption{Functional Coefficients used in the simulation study.}
    \label{fig:beta_simulation}
\end{figure}

The goal of the present Supplementary Section is to present the proposed methodology on a simulated case, showing how our methodology is able to capture time-varying sensitivity patterns, as well as to correctly define those areas of the domain where the impact of a specific input is statistically significant.
In order to perform the simulation study we explore 4 different scenarios: in the first two we employ a functional linear model of the form
\begin{equation}
    y(t)=\beta_1 (t) x_1 + \beta_2 (t) x_2 + \beta_3(t) x_3 + \epsilon(t),
\end{equation}
while in the second two we introduce an interaction between $x_1, x_2$ as follows
\begin{equation}
       y(t)=\beta_1 (t) x_1 + \beta_2 (t) x_2 + \beta_3(t) x_3 + \beta_{inter} (x_1 * x_2) + \epsilon(t).
\end{equation}
In both cases, for simplicity, $t\in[0,1]$. Coherently with the finite change framework described in the previous section, inputs $x_1, x_2, x_3$ are discrete ones, varying between ${0,1}$.
In order to successfully compute the different \emph{FCSI}s described in Sec. 2 we design a factorial experiment following a similar logic applied in \cite{borgonovo_sensitivity_2010,marangoni_sensitivity_2017}, thus having One-Factor-at-A-Time (OFAT) runs using as a reference level the $0$ one to compute Individual effects, and "reverse" OFAT runs for the Total effects.
We assume to have $10$ independent simulators, and thus we replicate the $8$ runs composing the factorial experiment $10$ times, thus having $80$ simulations.
In order to assess the domain-selective capabilities of the proposed methodology, we have built $\beta(t)$s with a specific structure.
More specifically, $\beta_1(t)$ is a parabola generated by the equation $y=-30 (t-0.5)^2 + 10$, $\beta_{inter}$ is instead a constant term $y=7$.
With respect to $\beta_2(t),\beta_3(t)$ and $\epsilon(t)$, we have created them using a B-spline basis of order $4$ composed of $10$ functions and equispaced knots. The $10$ coefficients of the basis expansion are then generated as follows
\begin{itemize}
    \item $\beta_2(t): \{U(8,10),U(8,10),U(8,10),N(0,1),N(0,1),N(0,1),N(0,1), \\ U(8,10),U(8,10),U(8,10) \}$ where $U(a,b)$ is the uniform distribution between $a,b$ and $N(a,b)$ is the normal distribution of mean $a$ and standard deviation $b$,
    \item $\beta_3(t) : N\left(\mathbf{0}, 0.1\, \mathbf{I}\right),$ where $\mathbf{X}$ is a 10-dimensional random vector, $\mathbf{0}$ is a 10-dimensional zero vector, and $\mathbf{I}$ is the $10$-dimensional identity matrix.
    \item $\epsilon(t) : N\left(\mathbf{0}, \sigma_\epsilon \, \mathbf{I}\right),$ where $\mathbf{X}$ is a 10-dimensional random vector, $\mathbf{0}$ is a 10-dimensional zero vector, and $\mathbf{I}$ is the $10$-dimensional identity matrix
\end{itemize} 
The shape of the different coefficients can be observed in Fig. \ref{fig:beta_simulation}.
The different coefficients represent different testing situations: $\beta_1(t)$ is a functional coefficient significant over the whole domain, $\beta_2(t)$ is significant only at the boundaries of the domain and $\beta_3(t)$ is never significant.

As previously mentioned, we propose 4 different scenarios:
\begin{itemize}
    \item Scenario 1: Interaction is absent, $\sigma_\epsilon=1$,  
    \item Scenario 2: Interaction is absent, $\sigma_\epsilon=5$,
    \item Scenario 3: Interaction is present, $\sigma_\epsilon=1$,
    \item Scenario 4: Interaction is present, $\sigma_\epsilon=5$.
\end{itemize}

We present the results in terms of the different FCSIs for the 10 different runs, as well as testing for the significance of the average FCSI over the runs as figures.

\begin{figure}
    \includegraphics[width=\linewidth]{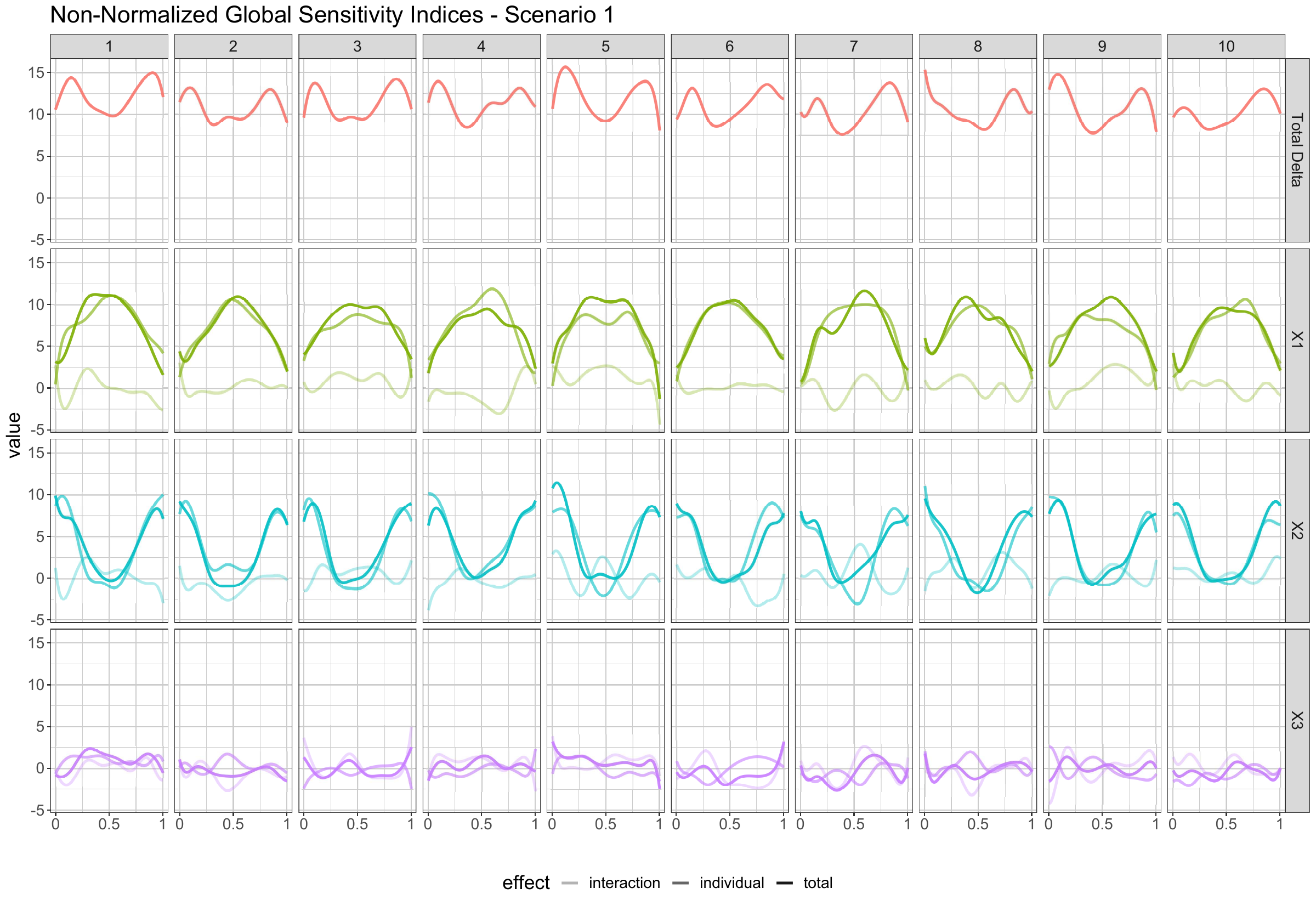}
    \caption{Functional Sensitivity indices for Scenario 1: In all the panels the y axis is the magnitude of the various non-normalized sensitivity indices. Different rows and different colors represent different factors, while we have different columns for different runs}
    \label{fig:sensitivities_scenario1}
\end{figure}
\begin{figure}
    \includegraphics[width=\linewidth]{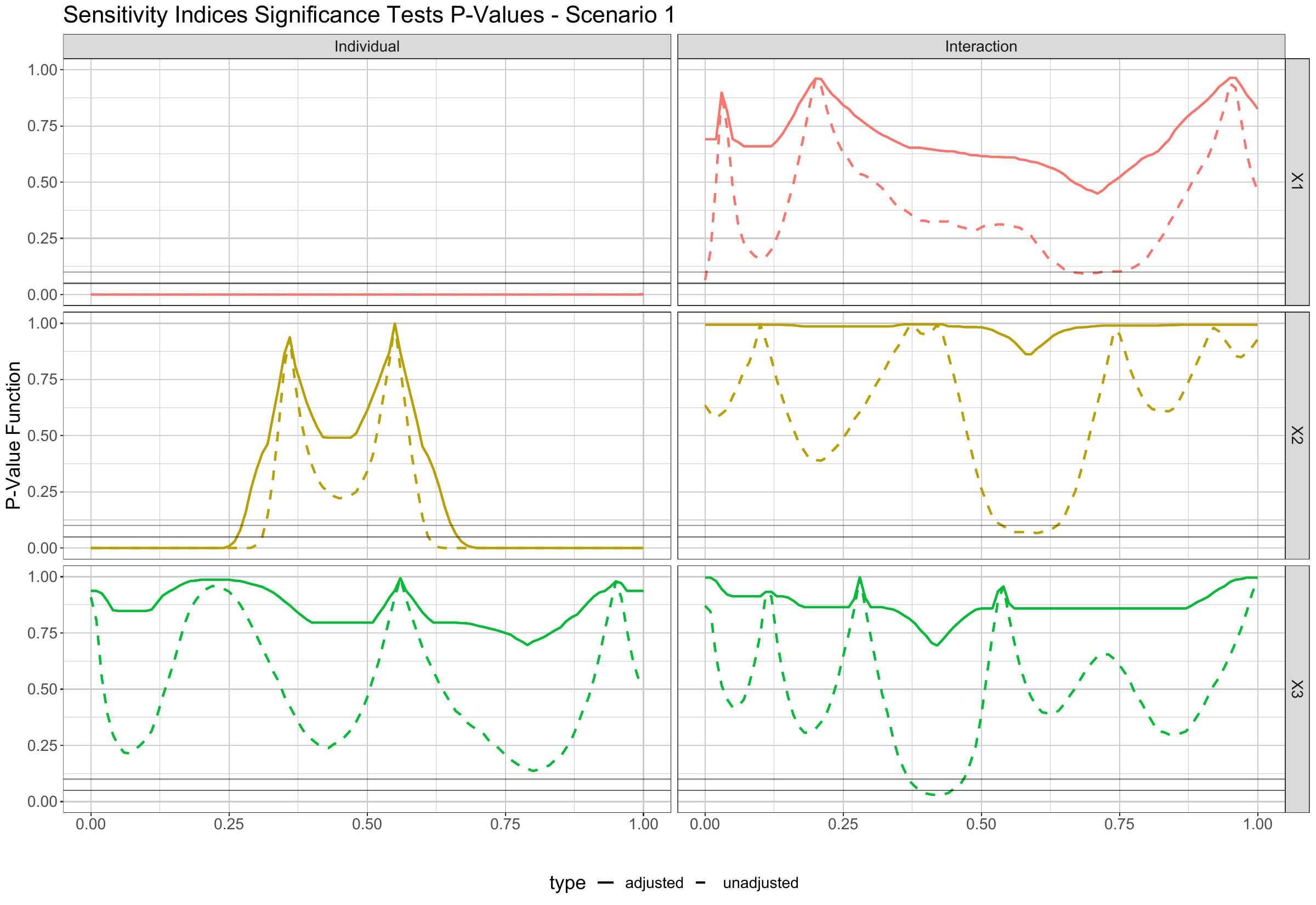}
    \caption{p-value functions for functional t-tests: Scenario 1. In all the panels the y axis shows the value of the adjusted (full line) and unadjusted (dotted line) p-value functions, from 0 to 1. Rows and colors denote different factors, while the two columns are for Individual and Interaction effects.}
    \label{fig:pvalues_scenario1}
\end{figure}
\begin{figure}
    \includegraphics[width=\linewidth]{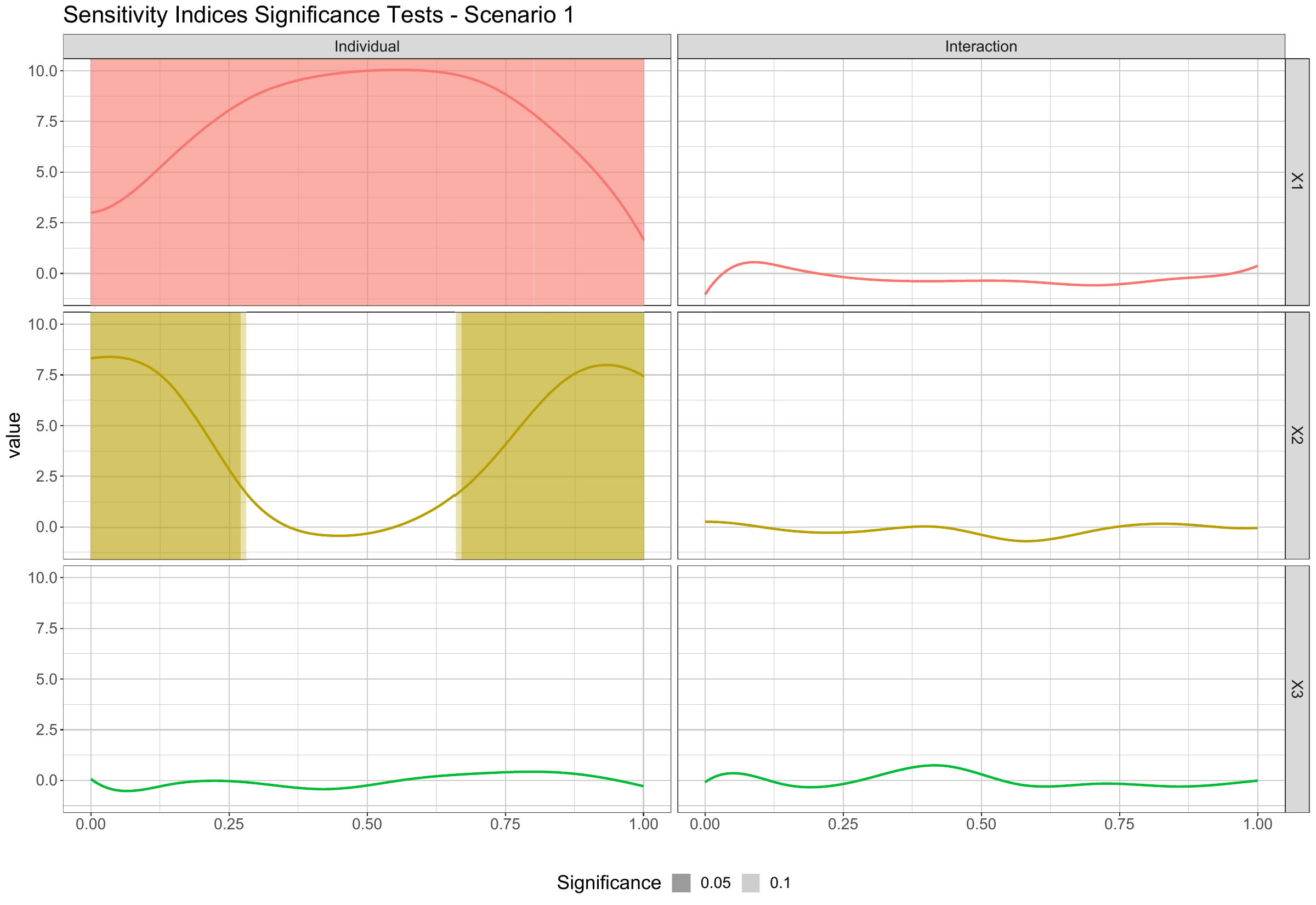}
    \caption{t-tests for Scenario 1. In all the panels, the y axis is the magnitude of the average sensitivity coefficient, different shading levels represent different significance levels, as denoted by the adjusted p-value functions (dark shading = 0.05 significance, light shading = 0.1 significance). Rows and colors denote different factors, while the two columns are for Individual and Interaction effects.}
    \label{fig:t_tests_scenario1}
\end{figure}

\begin{figure}
    \includegraphics[width=\linewidth]{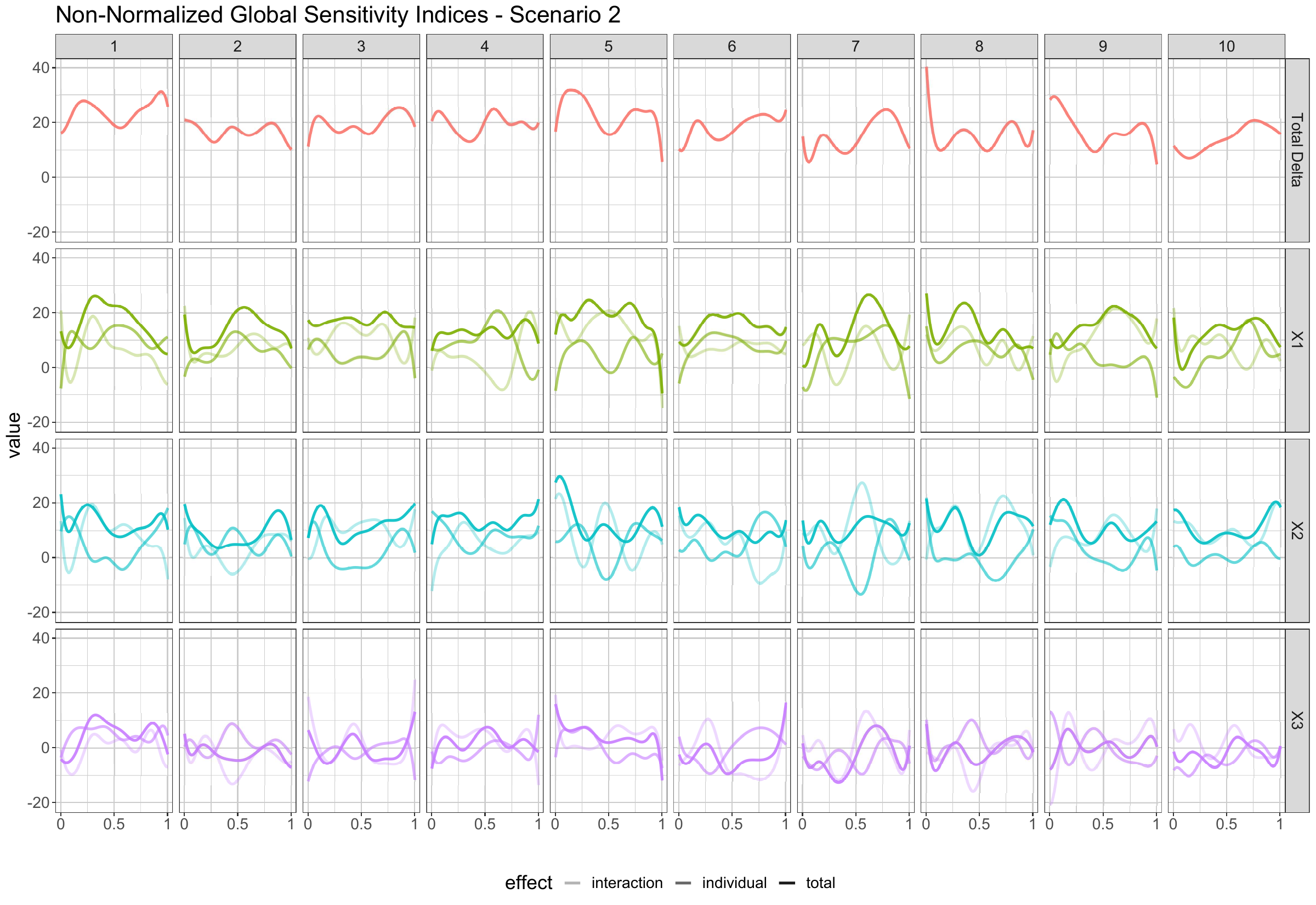}
    \caption{Functional Sensitivity indices for Scenario 2: In all the panels the y axis is the magnitude of the various non-normalized sensitivity indices. Different rows and different colors represent different factors, while we have different columns for different runs}
    \label{fig:sensitivities_scenario2}
\end{figure}
\begin{figure}
    \includegraphics[width=\linewidth]{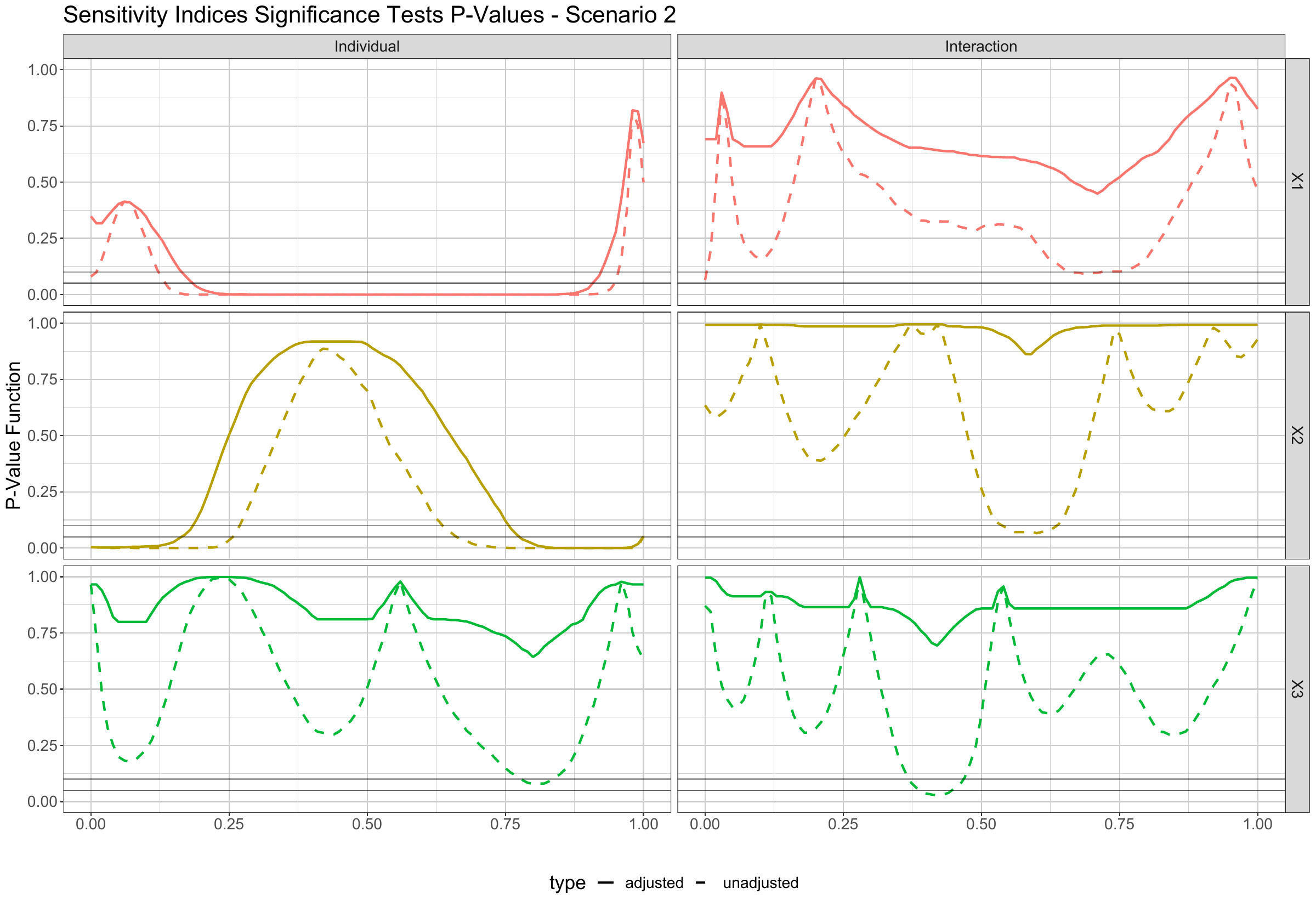}
    \caption{p-value functions for functional t-tests: Scenario 2. In all the panels the y axis shows the value of the adjusted (full line) and unadjusted (dotted line) p-value functions, from 0 to 1. Rows and colors denote different factors, while the two columns are for Individual and Interaction effects.}    \label{fig:pvalues_scenario2}
\end{figure}
\begin{figure}
    \includegraphics[width=\linewidth]{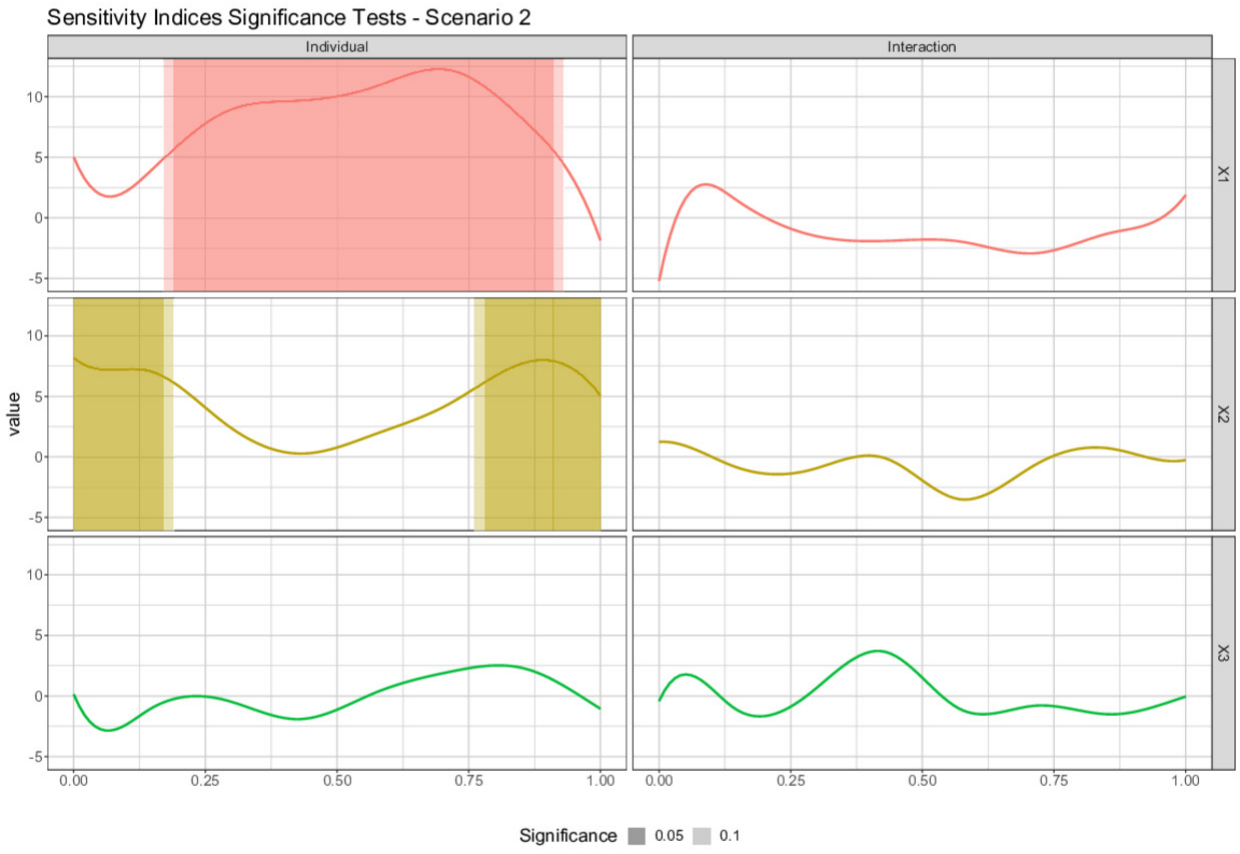}
    \caption{t-tests for Scenario 2. In all the panels, the y axis is the magnitude of the average sensitivity coefficient, different shading levels represent different significance levels, as denoted by the adjusted p-value functions (dark shading = 0.05 significance, light shading = 0.1 significance). Rows and colors denote different factors, while the two columns are for Individual and Interaction effects.}
    \label{fig:t_tests_scenario2}
\end{figure}

\begin{figure}
    \includegraphics[width=\linewidth]{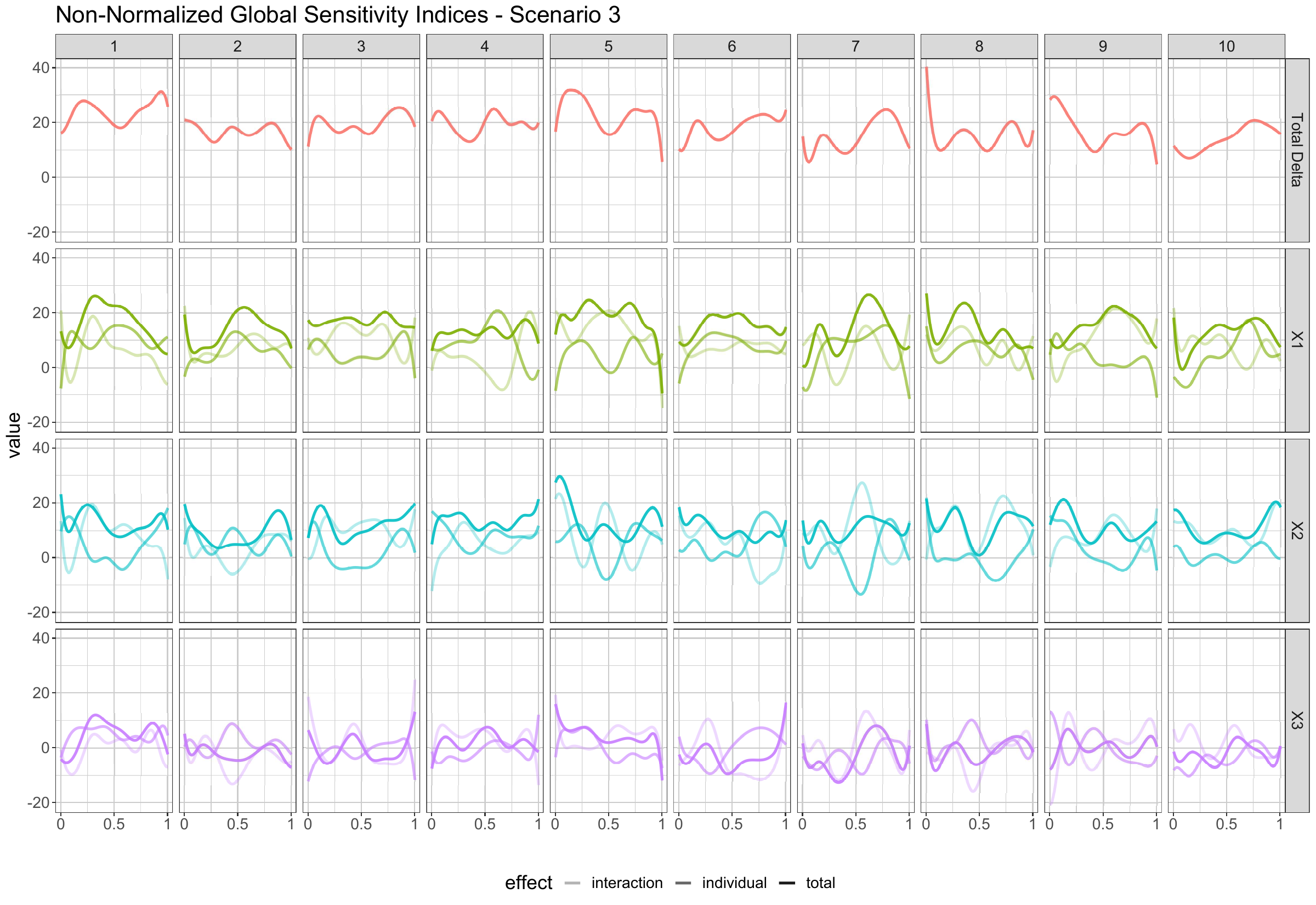}
    \caption{Functional Sensitivity indices for Scenario 3: In all the panels the y axis is the magnitude of the various non-normalized sensitivity indices. Different rows and different colors represent different factors, while we have different columns for different runs}
    \label{fig:sensitivities_scenario3}
\end{figure}
\begin{figure}
    \includegraphics[width=\linewidth]{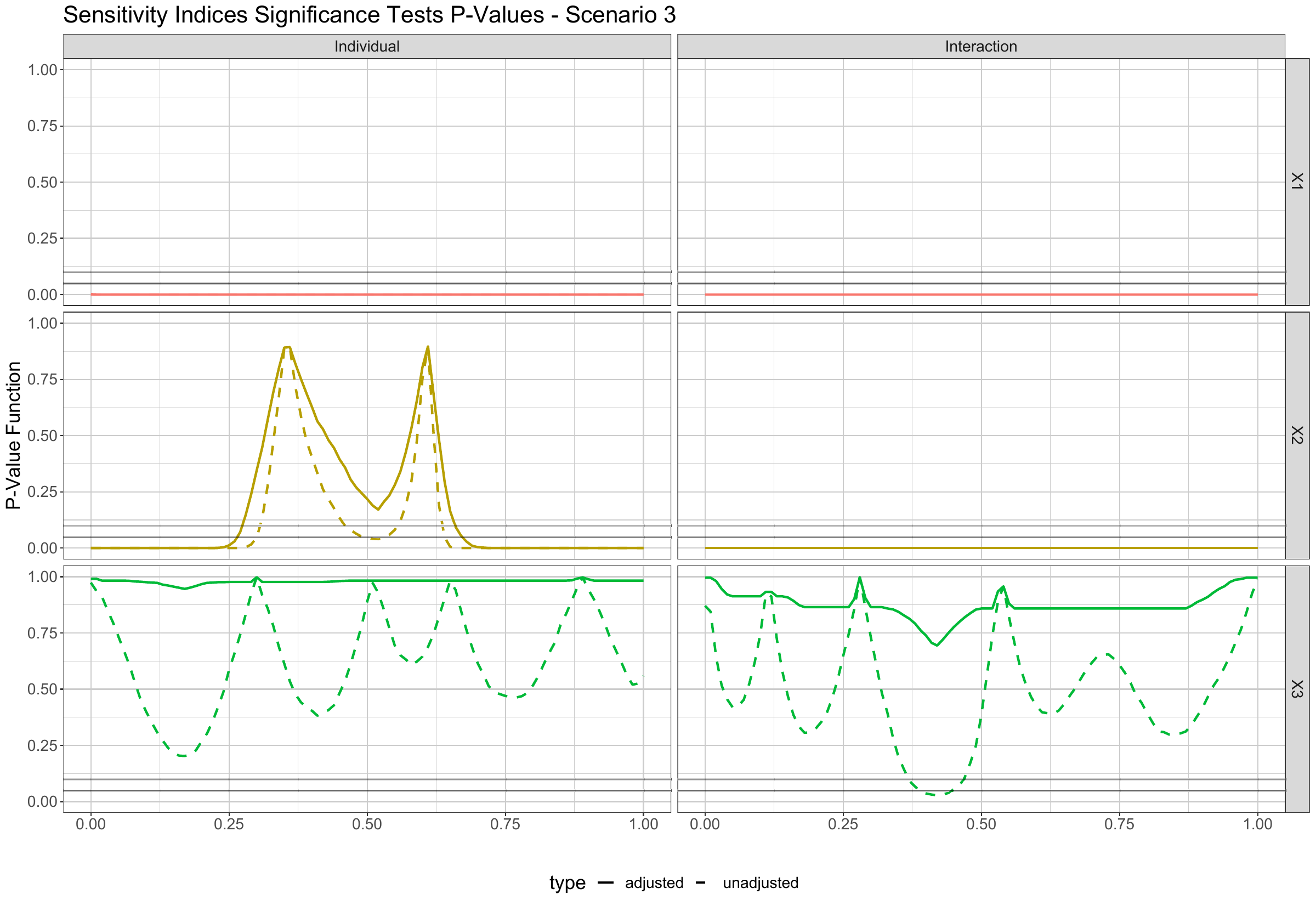}
    \caption{p-value functions for functional t-tests: Scenario 3. In all the panels the y axis shows the value of the adjusted (full line) and unadjusted (dotted line) p-value functions, from 0 to 1. Rows and colors denote different factors, while the two columns are for Individual and Interaction effects.}
    \label{fig:pvalues_scenario3}
\end{figure}
\begin{figure}
    \includegraphics[width=\linewidth]{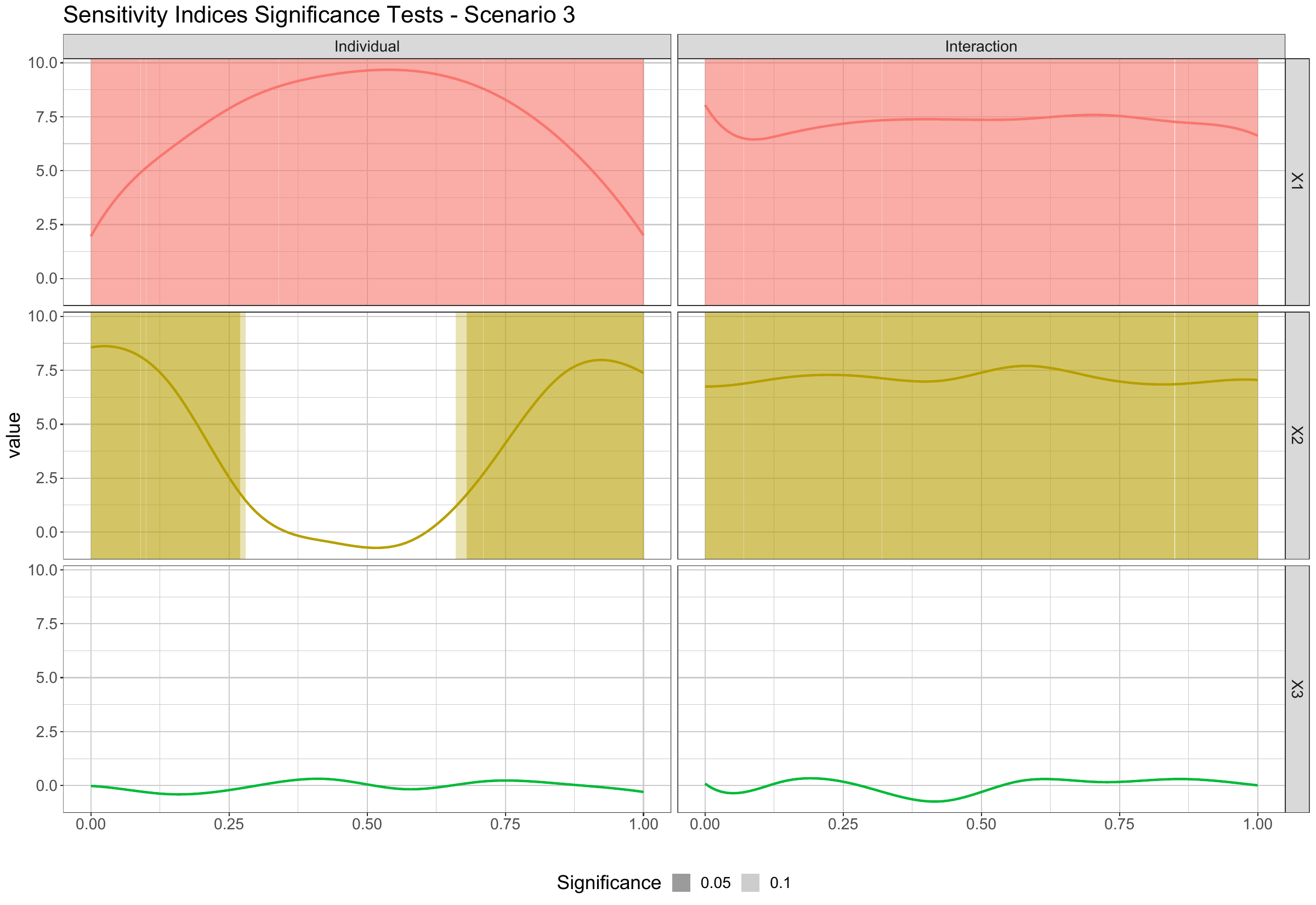}
    \caption{t-tests for Scenario 3. In all the panels, the y axis is the magnitude of the average sensitivity coefficient, different shading levels represent different significance levels, as denoted by the adjusted p-value functions (dark shading = 0.05 significance, light shading = 0.1 significance). Rows and colors denote different factors, while the two columns are for Individual and Interaction effects.}
    \label{fig:t_tests_scenario3}
\end{figure}

 \begin{figure}
    \includegraphics[width=\linewidth]{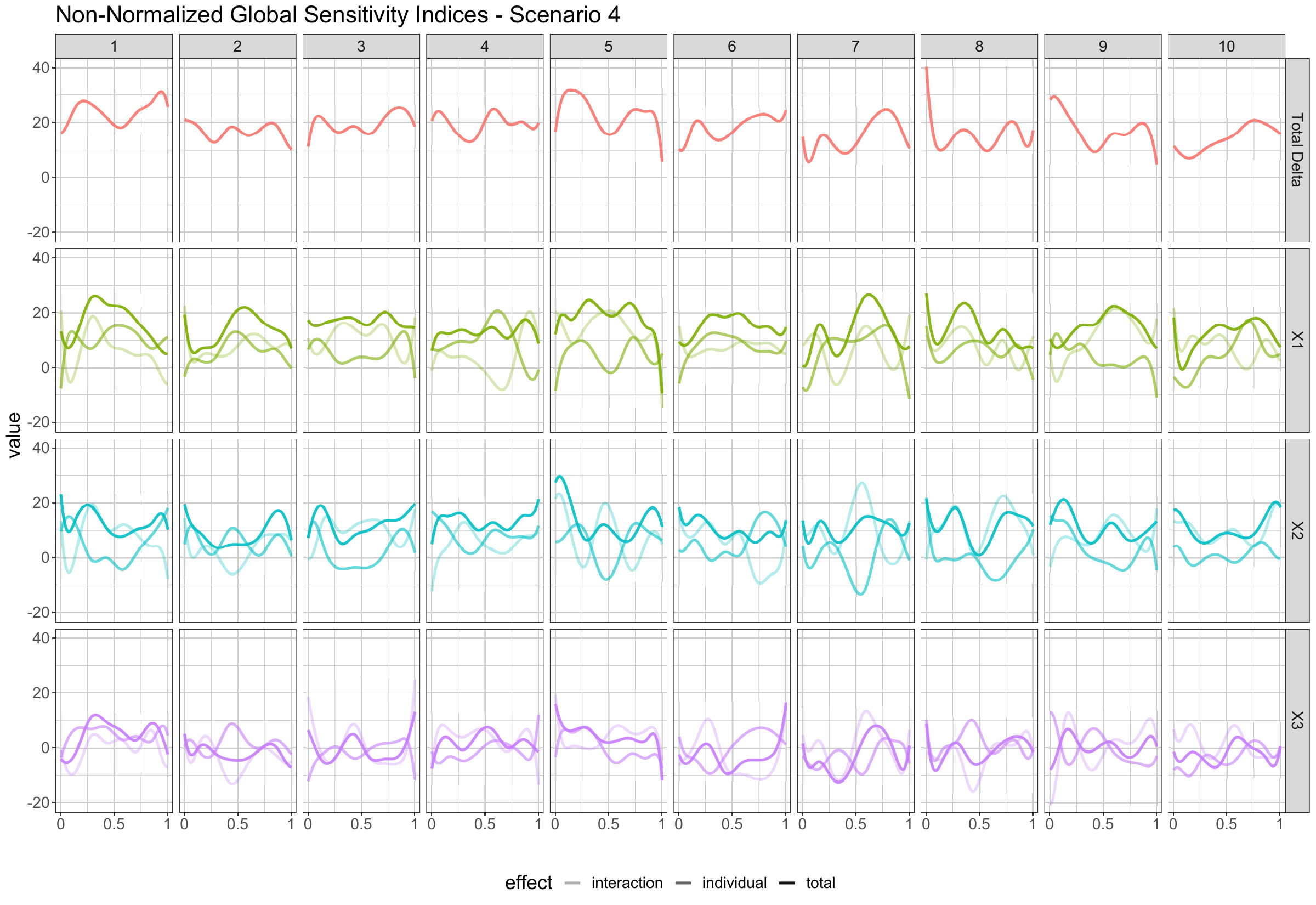}
    \caption{Functional Sensitivity indices for Scenario 4: In all the panels the y axis is the magnitude of the various non-normalized sensitivity indices. Different rows and different colors represent different factors, while we have different columns for different runs}
    \label{fig:sensitivities_scenario4}
\end{figure}
\begin{figure}
    \includegraphics[width=\linewidth]{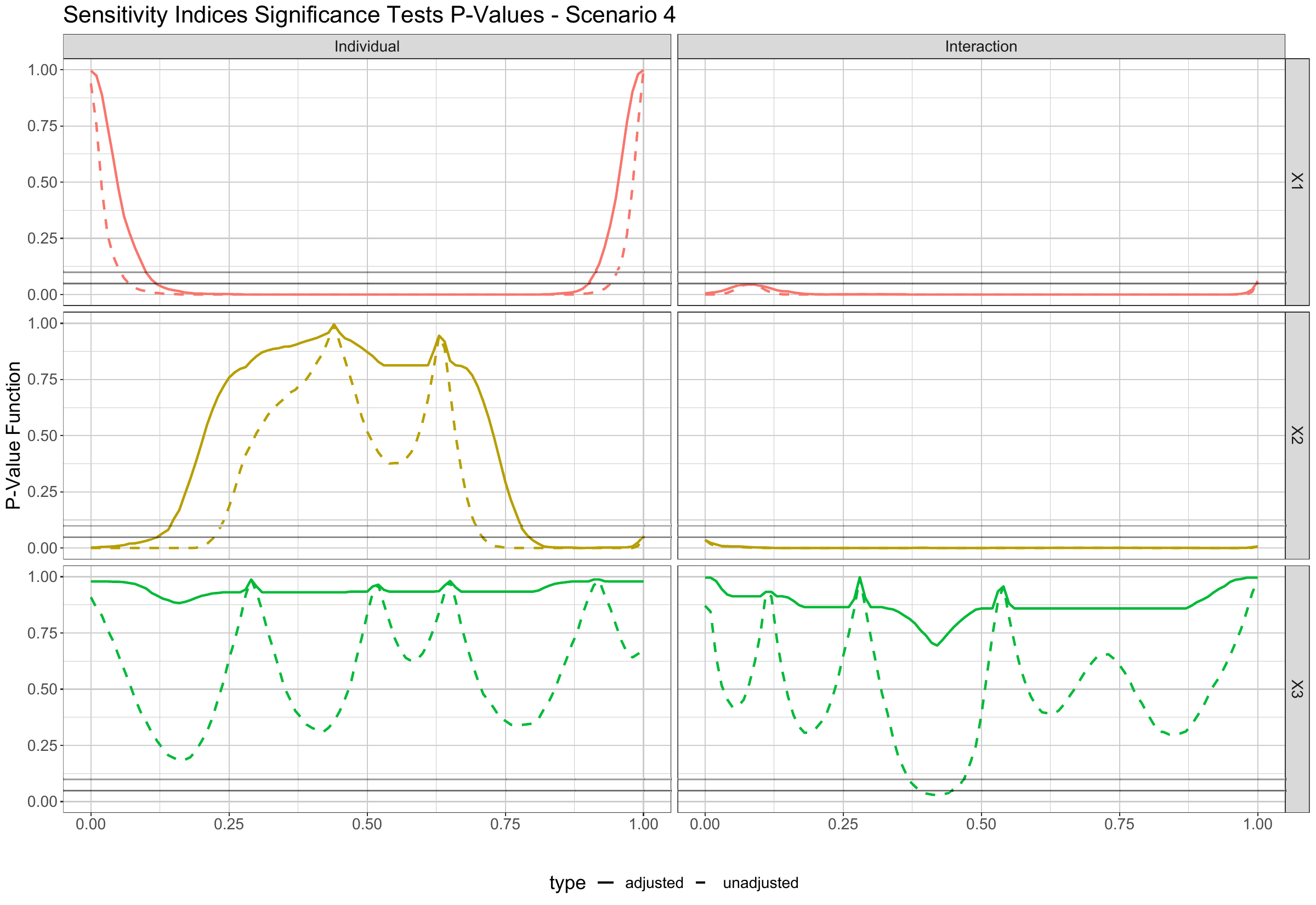}
    \caption{p-value functions for functional t-tests: Scenario 4. In all the panels the y axis shows the value of the adjusted (full line) and unadjusted (dotted line) p-value functions, from 0 to 1. Rows and colors denote different factors, while the two columns are for Individual and Interaction effects.}
    \label{fig:pvalues_scenario4}
\end{figure}
\begin{figure}
    \includegraphics[width=\linewidth]{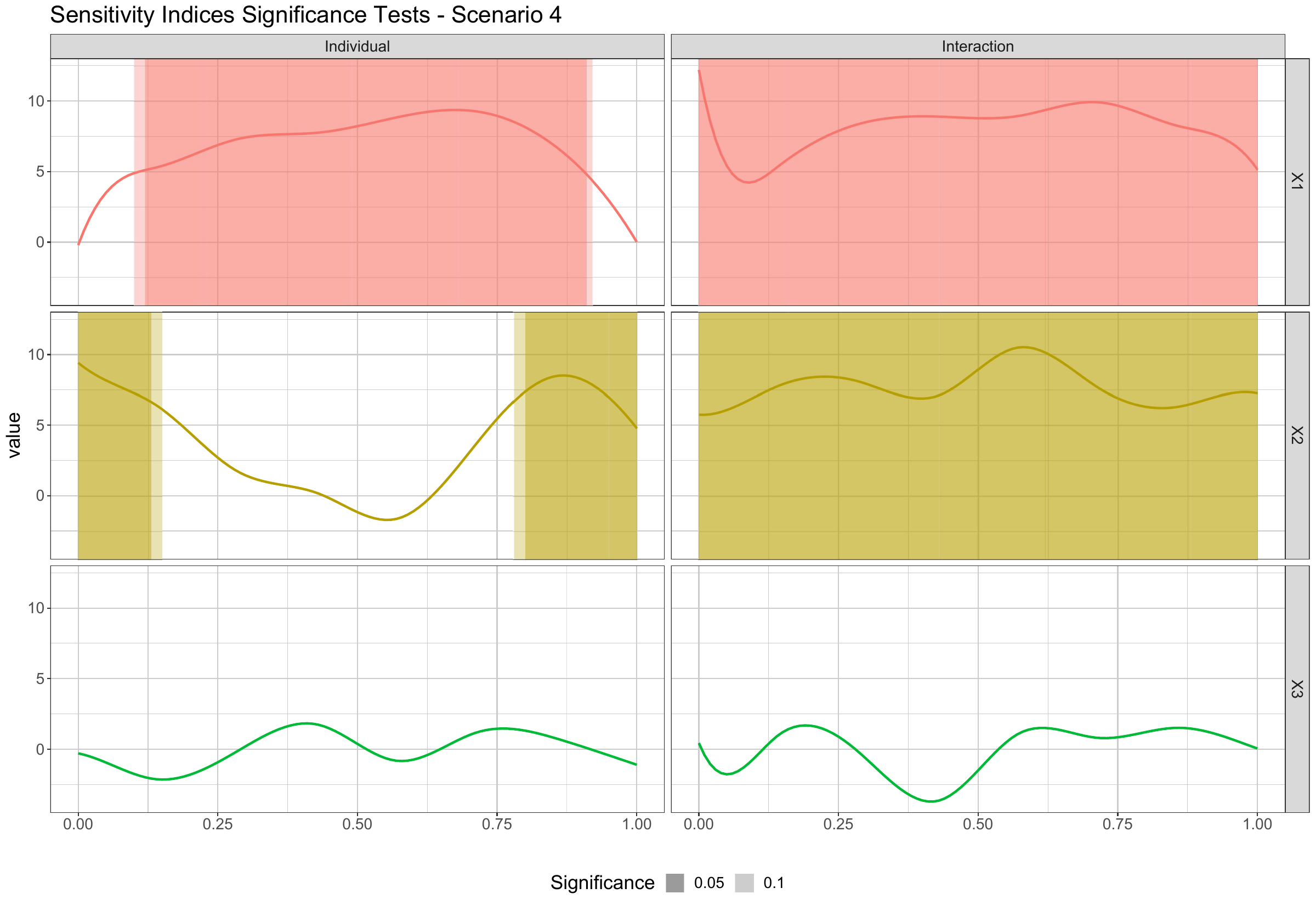}
    \caption{t-tests for Scenario 4. In all the panels, the y axis is the magnitude of the average sensitivity coefficient, different shading levels represent different significance levels, as denoted by the adjusted p-value functions (dark shading = 0.05 significance, light shading = 0.1 significance). Rows and colors denote different factors, while the two columns are for Individual and Interaction effects.}
    \label{fig:t_tests_scenario4}
\end{figure}

Starting from Scenario 1, whose Sensitivity indices are depicted in Fig. \ref{fig:sensitivities_scenario1}, p-value functions in Fig. \ref{fig:pvalues_scenario1} and t-tests in Fig. \ref{fig:t_tests_scenario1}, we can immediately observe how the proposed methodology is not only able to provide a clear picture in terms of sensitivities (The shapes of the FCSIs are coherent among each other), but also, by using the regression and testing step, it is able to define correctly the shapes as well as the significance patterns. Namely, the shapes of the functions are correctly identified and we see significance over the whole domain for the FCSI of $x_1$, significance concentrated at the boundaries of the domain for $x_2$, and no significance for $x_3$. Moreover, and correctly, Interaction effects are not identified.

With respect to Scenario 2, whose Sensitivity indices are depicted in Fig. \ref{fig:sensitivities_scenario2}, p-value functions in Fig. \ref{fig:pvalues_scenario2} and t-tests in Fig. \ref{fig:t_tests_scenario2}, the situation is similar as Scenario 1, showing that our methodology is robust also to high-noise situations.
Moreover, in this case it appears immediately evident how our regression and testing approch proves to be fundamental, as the situation described by the different sensitivities indices is way less clear due to noise, and our method contributes in clarifying.

The same conclusions can be extracted from Scenario 3 and 4, where apart from the good properties found in analysing the previous two scenarios, we are also able to correctly define and assess the significance of interaction effects, and discriminate where interactions are not present (namely for $x_3$)

\section{Bibliography}

\bibliographystyle{apa}
\bibliography{references,references_old}